 \documentclass[final,5p,times,twocolumn,authoryear]{elsarticle}

\usepackage{amssymb}
\usepackage{lipsum}
\usepackage{xcolor}
\usepackage{float}
\usepackage{stfloats} 

\usepackage[]{hyperref}
\hypersetup{
setpagesize=false,
 colorlinks=true,
 linkcolor=blue,
 citecolor=blue,
 filecolor=blue,
 urlcolor=blue,
}

\journal{High Energy Astrophysics}

\begin{document}

\begin{frontmatter}



\title{
%
Revisiting early afterglows of gamma-ray bursts with finite-thickness ejecta:
Implications from XRF~080330 and GRB~080710
}



\author[1]{Kaori~Obayashi}
\author[1,2]{Ryo~Yamazaki}
\author[3]{Yo~Kusafuka}
\author[3]{Katsuaki~Asano}
\affiliation[1]{organization={Department of Physical Sciences, Aoyama Gakuin University},
            addressline={5-10-1 Fuchinobe}, 
            city={Sagamihara},
            postcode={252-5258}, 
            state={Kanagawa},
            country={Japan}}
\affiliation[2]{organization={Institute of Laser Engineering, Osaka University},
            addressline={2-6, Yamadaoka}, 
            city={Suita},
            postcode={565-0871}, 
            state={Osaka},
            country={Japan}}
\affiliation[3]{organization={Institute for Cosmic Ray Research, The University of Tokyo},
            addressline={5-1-5 Kashiwanoha}, 
            city={Kashiwa},
            postcode={277-8582}, 
            state={Chiba},
            country={Japan}}

\begin{abstract}

We revisit the physical origin of the achromatic peaks and breaks observed several thousand seconds after the burst in the multi-wavelength afterglows of XRF~080330 and GRB~080710. Using a numerical afterglow model that consistently incorporates finite ejecta thickness and a generalized external density profile, we perform Bayesian inference to estimate model parameters describing these events.
Our analysis shows that the gradual rise and achromatic temporal features in both events are more naturally explained by jet dynamical evolution with finite shell thickness rather than by off-axis viewing effects. The inferred initial radial width of the ejecta is of order $10^{13}$~cm for both bursts, implying a central engine activity timescale significantly longer than that suggested by the prompt gamma-ray duration alone.  
Taken together, these results demonstrate that early afterglow light curves are strongly influenced by transition dynamics when finite ejecta thickness is properly taken into account, thereby providing a physical link between the prompt and afterglow phases 
and highlighting limitations of simply applying the thin-shell approximation when interpreting early-time afterglows.
Furthermore, Bayesian model comparison strongly favors a generalized circumburst density profile over the canonical uniform or steady-wind models, suggesting that fixing the external density structure to idealized profiles a priori may obscure crucial information about the progenitor's pre-burst activity.

\end{abstract}




\end{frontmatter}




\section{Introduction}
\label{introduction}

Gamma-ray bursts (GRBs) are among the most 
electromagnetically luminous explosive phenomena and 
are widely interpreted as arising from
relativistic jets produced by a compact central engine 
\citep[e.g.][]{piran1999, kumar2015, gotz2018}.
After the prompt gamma-ray emission, 
broadband afterglows are commonly observed and are most likely
produced by synchrotron radiation from non-thermal electrons 
accelerated at relativistic external shocks as the jet interacts with the circumburst medium (CBM) \citep{sari1998, sari1999a}.
Since the launch of the \textit{Neil Gehrels Swift Observatory} in 2004 \citep{gehrels2004}, early afterglow observations beginning within minutes of the burst trigger have become routinely available.
These observations have revealed a variety of light-curve behaviors in various observed frequency bands
that are difficult to reconcile with predictions of the standard afterglow model established prior to the \textit{Swift} era.
For example, 
early X-ray afterglows often show a canonical behaviour
--- an initial steep decay, followed by a shallow-decay phase and then a normal decay phase 
\citep[e.g.][]{nousek2006, zhang2006, willingale2007, liang2007, zhang2007} ---
which cannot be straightforwardly explained by a single 
external forward-shock model.
Very early optical afterglows 
also exhibit remarkable diversity, with luminosities spanning three to four orders of magnitude at
$ \sim 500\,\mathrm{s}$ after the GRB trigger \citep[e.g.,][]{kann2010, gao2015, wang2013, liang2013, panaitescu2011}. 
A significant fraction of events are optically faint 
\citep[e.g.,][]{roming2006}, 
challenging standard expectations for prominent reverse-shock emission
\citep{sari1995, kobayashi2000, kobayashi2000b}.
Moreover, some observational results indicate that early X-ray and optical afterglows do not necessarily share a common emission site
\citep[e.g.,][]{panaitescu2006, sato2007, urata2007, oates2007, liang2008, liang2009, liang2015, guidorzi2025},
suggesting distinct physical origins
\citep{ghisellini2007, yamazaki2009}.

In this context,
the \textit{Einstein Probe} satellite, 
with its high sensitivity in the soft X-ray band \citep{yuan2022},
is expected to open a new window on GRB research.
In particular, the X-ray flashes (XRFs) 
\citep{gotthelf1996,hamilton1996,strohmayer1998,heise2001,arefiev2003,barraud2003,intzand2025}
are of great interest, 
but their origin remains uncertain \citep{yuan2025}.
Observationally, 
XRFs are characterized by low spectral peak energies and weak gamma-ray emission, and
are generally regarded as part of a continuous phenomenological sequence with classical GRBs \citep{sakamoto2005,sakamoto2008,dalessio2006}.
Understanding the origin of XRFs could yield important insights into the 
nature of classical GRBs.
Various theoretical models of XRFs have been proposed:
high-redshift GRBs \citep{heise2001,barraud2003},
top-hat jets viewed off-axis \citep{ioka2001,yamazaki2002, yamazaki2003b, yamazaki2004a},
wide opening-angle jets \citep{lamb2005,donaghy2006},
internal shocks with a small contrast in high Lorentz factors \citep{daigne2003},
failed GRBs or dirty fireballs \citep{dermer1999, huang2002, dermer2004},
photosphere-dominated fireballs \citep{meszaros2002, ramirez2002, drenkhahn2002},
structured or inhomogeneous jets \citep{rossi2002, zhang2002,zhang2004,yamazaki2004b,toma2005},
peripheral emissions from collapsar jets \citep{zhang2004},
and off-axis cannonballs \citep{dar2004}.
As the number of detected XRFs increases rapidly in the
\textit{Einstein Probe} era \citep{gao2025},
early afterglows of XRFs and GRBs will provide a stringent test of these competing theoretical models.

Several events show early optical and near-infrared (NIR) 
afterglows with  simultaneous brightness maximum across multiple bands at $t \sim 10^{3-4}$~s, forming so-called achromatic peaks
\citep{liang2010, liang2013, ror2025}.
Such late-time achromatic peaks are 
challenging to accommodate within the standard afterglow framework, 
where emission at $\gtrsim 10^2$~s is attributed to radiation from the decelerating external shock, and the evolution of characteristic synchrotron frequencies predicts chromatic peaks appearing at different times in different observing bands \citep{sari1998, sari1999a, sari1999b}.
XRF~080330 and GRB~080710 are
representative events exhibiting achromatic peaks at $t \sim 10^3$~s \citep{kruhler2009b, guidorzi2009}.
In the observation and analysis papers \citep{kruhler2009b, guidorzi2009},
both events were qualitatively interpreted as off-axis afterglows 
\citep[e.g.][]{granot2002, totani2002, ramirez2005, granot2005, lamb2017, ryan2020}.
However, if XRFs are primarily off-axis classical GRBs, explaining the achromatic peaks in both XRF~080330 and GRB~080710 as off-axis afterglows would imply an XRF-like prompt emission in both events.
This is difficult to reconcile with the fact that GRB~080710 is observationally classified as a classical GRB.

While off-axis viewing remains a plausible explanation, 
other physical ingredients can also generate achromatic 
temporal features.
For GRB~080710, 
\citet{obayashi2024} has shown
that the off-axis scenario is not necessarily the most favored interpretation and instead suggested
an on-axis origin driven by ejecta dynamics.
Another possibility is that 
a non-uniform CBM and/or finite-thickness ejecta can modify the shock dynamics, 
thereby affecting the temporal evolution of the early afterglow light curves in an achromatic manner \citep{yi2013, kobayashi2000}.
In addition, 
recent numerical studies have shown that thick-shell ejecta can 
lead to shock evolution
that differs qualitatively from that predicted by the thin-shell approximation, producing distinct early afterglow behavior
\citep{kusafuka2023, kusafuka2025a, kusafuka2025b}.

In this work, we revisit the early afterglows of XRF~080330 and 
GRB~080710, focusing on the effects of finite ejecta thickness and a generalized CBM density profile.
Our goal is to quantify how early afterglow light curves constrain  
the initial conditions of the ejecta, the circumburst environment and microphysical parameters within an external shock model that accounts for finite ejecta thickness and a power-law CBM density profile.
In particular, because early-afterglow modeling inevitably involves multiple candidate physical ingredients and strong parameter degeneracies, we adopt a Bayesian framework that enables a direct comparison of posterior distributions and Bayesian evidence, thereby assessing which ingredients are truly required (e.g., off-axis geometry, finite thickness, or CBM structure) within a unified statistical approach. By analyzing the two contrasting events that share similar achromatic temporal features but have markedly different prompt-emission properties \citep{kruhler2009b, guidorzi2009},
we aim to extract robust implications for early afterglow physics and to gain insight into progenitor environments and jet-formation processes.
%

This paper is organized as follows.
In Section~\ref{sec:thickshell}, we introduce the model of afterglow emission produced by finite-thickness ejecta.
In Section~\ref{sec:codebaysian}, we explain the numerical calculation and the model-evaluation method.
We use \texttt{Magglow} \citep{kusafuka202511}, 
which is well suited to our purpose.
In Section~\ref{sec:result}, we present the results,
and Section~\ref{sec:discussion} is devoted to discussion.
Throughout this paper, we adopt a concordance cosmology with
$H_0 = 70~\mathrm{km\,s^{-1}\,Mpc^{-1}}$,
$\Omega_\Lambda=0.7$ and $\Omega_M = 0.3$.
The redshift of XRF~080330 ($z=1.51$)
corresponds to the luminosity distance 
$d_L = 3.4 \times 10^{28}\,\mathrm{cm}$,
while the redshift of GRB~080710 ($z=0.845$) corresponds to
$d_L = 1.7 \times 10^{28}\,\mathrm{cm}$.

\section{Dynamics of ejecta with finite thickness and resulting afterglow emission}
\label{sec:thickshell}

In this section, we outline the forward-shock dynamics modified by the finite ejecta thickness and a non-uniform CBM, as well as the resulting afterglow emission, both of which are expected to influence the early afterglow evolution 
\citep[e.g.,][]{sari1995,sari1997thick,yi2013,gill2023,kusafuka202511}.
Following \citet{kusafuka202511}, 
we adopt a coordinate system in which the central engine that launches the jet is located at the origin, $R = 0$. 
Then, the density structure of the CBM is expressed as a function of radius $R$ as 
\begin{equation}
  n_\mathrm{CBM}(R) = A(n_0, k) R^{-k}~~,
\end{equation}
where we set
$A(n_0, k) = n_0 (3\times 10^{35})^{k/2}$,
and $n_0$ and $k$ determine the normalization and power-law index,
respectively\footnote{
Note that the parameter $n_0$ has units of 
$[\mathrm{cm^{k-3}}]$.
For example, when $k=0$, this corresponds to a uniform interstellar medium (ISM) model with
$n_\mathrm{CBM}=n_0$.
When $k=1$, the surrounding environment is formed by a non-steady stellar wind, and the density profile becomes $n_\mathrm{CBM}(R)=\sqrt{30}\times(n_0/1~\mathrm{cm^{-2}})(R/10^{17}~\mathrm{cm})^{-1}~\mathrm{cm^{-3}}$.
Finally, for $k=2$, 
we obtain $n_\mathrm{CBM}(R)=30\times (n_0/1~\mathrm{cm^{-1}})(R/10^{17}~\mathrm{cm})^{-2}~\mathrm{cm^{-3}}$
as a stationary stellar-wind profile.}.
The ejecta are characterized by 
the isotropic-equivalent kinetic energy $E_0~[\mathrm{erg}]$,
the initial Lorentz factor $\Gamma_0$,
and the initial radial width of the ejecta $\Delta_0~[\mathrm{cm}]$.
Although we adopt the dynamical framework of \citet{kusafuka202511}, 
we neglect the effects of magnetic fields in the ejecta for simplicity.
We also ignore the sideways expansion of the jet since we mainly 
focus on the early phase of the afterglows.

In the widely used thin-shell approximation for GRB afterglow models,
the jet starts to decelerate at the deceleration radius 
$R_\gamma = l_S \Gamma_0^{-2/(3-k)}$,
at which the energy the swept-up ambient material have gained
becomes comparable to the initial explosion energy.
Here, $l_S$ is defined for $0<k<3$ as:
\begin{equation}
    l_S \equiv \left(\frac{(3-k)E_0 }{4\pi m_\mathrm{p} c^2 A(n_0, k)}\right)^{\frac{1}{3-k}}.
\end{equation}
Although the model used in this study allows for an arbitrary viewing angle,
for simplicity we assume an on-axis observer when evaluating the observer time and characteristic timescales in the following analytic evaluation.
The photons emitted from the deceleration radius $R_\gamma$
are detected at the observer time $T_\gamma$, which is given by
\begin{eqnarray}
 T_\gamma
  &\sim& \frac{(1+z)R_\gamma}{2c\Gamma_0^2} \nonumber \\
  &\sim& (1+z)\frac{l_S}{2c}\Gamma_0^{-\frac{8-2k}{3-k}}~~,
  \label{eq:Tdec}
\end{eqnarray}
where $z$ is the redshift of the source \citep{yi2013}.
This timescale is often called the deceleration time or afterglow onset time.
\citet{kusafuka2025a, kusafuka2025b}, however, 
studied the deceleration of an ejecta shell with finite thickness, 
and demonstrated that 
the onset of  deceleration following 
the Blandford-McKee (BM) scaling \citep{blandfordmckee1976}
does not necessarily coincide with $T_\gamma$ but occurs at a later time.

The dynamical evolution of the jet starts with
 the free-expansion (coasting) phase,
in which the Lorentz factor of the ejecta is approximately constant.
This phase is followed by two subsequent stages:
the transition phase and the adiabatic expansion phase.
The former appears as a consequence of the shell thickness.
The latter is described by the conventional BM scaling. 
The Lorentz factor of the forward shock in the free-expansion 
and transition phases is given by \citep{kusafuka2025a}:
\begin{eqnarray}
\label{eq:GammaFS_trans_origin}
    \Gamma_\mathrm{FS}
  &=& \frac{\Gamma_0}{\left[1+2\Gamma_0\sqrt{\frac{n_\mathrm{CBM}}{n_\mathrm{ej}}}\right]^{1/2}}
  \nonumber \\
  &=& \Gamma_0\left(1 + 2\Gamma_0^2\sqrt{\frac{(3-k)\Delta_0}{l_S^{3-k}}}R^{\frac{2-k}{2}}\right)^{-\frac{1}{2}}~~.
\end{eqnarray}
Hence, if $k=2$, then $\Gamma_\mathrm{FS}$ is constant with $R$.
In the remainder of this section, we consider the case of $k\neq2$.
For  $n_\mathrm{CBM}\ll n_\mathrm{ej}/\Gamma_0^2$, 
we have
\begin{equation}
\label{eq:GammaFS_free}
    \Gamma_\mathrm{FS, free}(R) \sim \Gamma_0~~.
\end{equation}
On the other hand, for $n_\mathrm{CBM} \gg n_\mathrm{ej}/\Gamma_0^2$, 
Eq.~(\ref{eq:GammaFS_trans_origin}) is approximated as
\begin{equation}
\label{eq:GammaFS_trans}
    \Gamma_\mathrm{FS, trans}(R)\sim \left(\frac{l_S^{3-k}}{4(3-k)\Delta_0}\right)^{1/4}R^{-\frac{2-k}{4}}~~.
\end{equation}
The observer time, $T_\mathrm{tr}$, at which this transition phase begins
can be evaluated with the radius $R_\mathrm{tr}$ that satisfies $n_\mathrm{CBM}(R_\mathrm{tr})\sim n_\mathrm{ej}(R_\mathrm{tr})/4\Gamma_0^2$.
For an on-axis observer, we get
\begin{eqnarray}
\label{eq:Ttra}
    T_\mathrm{tr}
    &\approx& (1+z)\frac{R_\mathrm{tr}}{2c\Gamma_0^2} \nonumber \\
    &\approx& \frac{1+z}{2c\Gamma_0^2}\left(\frac{l_S^{3-k}}{4(3-k)\Delta_0 \Gamma_0^4}\right)^{\frac{1}{2-k}}~~.
\end{eqnarray}

In the early epoch of the transition phase, 
the reverse shock propagates towards the inner edge of the shell.
After the reverse shock has crossed the ejecta at a radius $R_\Delta$, 
a rarefaction wave propagates toward the forward shock front. 
The radius  $R_\Delta$ is given by 
\begin{eqnarray}
\label{eq:RDelta}
    R_\Delta &= C_\mathrm{RS}^{\frac{2}{4-k}}\left(\frac{1}{3-k}\right)^{\frac{1}{4-k}}l_S^{\frac{3-k}{4-k}}\Delta_0^{\frac{1}{4-k}}~~,
\end{eqnarray}
where $C_{RS}$ is a nondimensional constant of order unity that depends on the reverse-shock regime \citep{sari1995, yi2013}.
This is converted to the observer time for an on-axis observer, 
$T_\Delta$, which is obtained from Eqs.~(\ref{eq:GammaFS_trans}) and (\ref{eq:RDelta}) as
\begin{eqnarray}
    T_\Delta
    &\sim& \frac{(1+z)R_\Delta}{2c\Gamma_\mathrm{FS,trans}(R_\Delta)^2}
     \nonumber \\
    &\sim& (1+z)C_\mathrm{RS}\frac{\Delta_0}{c}~~.
\end{eqnarray}

Once the rarefaction wave catches up with the shock front, 
the information about the initial shell thickness disappears, 
and the dynamics of the forward shock approaches 
the BM self-similar solution. 
The timing of this transition to the BM solution, 
$T_\mathrm{BM}$, can be significantly later than the deceleration time $T_\gamma$ estimated 
in the conventional framework \citep{kusafuka2025a, kusafuka2025b}.
We estimate the time interval $\Delta T_\mathrm{RW}$, measured by an on-axis observer, for the rarefaction wave to propagate across the shell and reach the forward shock.
In the lab frame, the shell has a thickness $\Delta_0$, while in the comoving frame of the ejecta, its thickness is $\Delta_0'=\Gamma_s\Delta_0$, where $\Gamma_s$ is the Lorentz factor of the shocked ejecta measured in the rest frame of the central engine.
In the comoving frame, the time for the rarefaction wave to traverse this thickness $\Delta_0'$ at the sound speed $c_s=c/\sqrt{3}$ is $\Delta t_\mathrm{RW}'=\Delta_0'/c_s=\Gamma_s\Delta_0/c$.
Transforming this into the lab-frame time gives 
$\Delta t_\mathrm{RW}=\Gamma_s\Delta t_\mathrm{RW}'=\Gamma_s^2\Delta_0/c$.
Finally, converting to the observer time yields 
$\Delta T_\mathrm{RW}\sim (1+z)\Delta t_\mathrm{RW}/(2\Gamma_s^2) 
\sim (1+z)\sqrt{3}\Delta_0/c$.
Therefore, the end time of the transition phase, 
$T_\mathrm{BM}$, is estimated as
\begin{eqnarray}
\label{eq:TBM}
    T_\mathrm{BM} 
    &=& T_\Delta+\Delta T_{\rm RW} \nonumber \\
    &\sim& (1+z)(C_\mathrm{RS}+\sqrt{3})\frac{\Delta_0}{c}~~.
\end{eqnarray}
The Lorentz factor of the forward shock after $T_\mathrm{BM}$ follows the BM scaling:
\begin{equation}
\label{eq:GammaFS_wind_1}
    \Gamma_\mathrm{FS, dec}(R) 
  = l_S^{\frac{3-k}{2}} R^{-\frac{k-3}{2}}.
\end{equation}

So far, we have neglected the expansion of the shell, so that
the thickness $\Delta_0$ is constant.
Since a velocity dispersion exists in the shell, the effect of shell spreading sometimes becomes important.
The radius at which shell spreading becomes significant is $R_\mathrm{sp}\sim \Delta_0\Gamma_0^2$.
This radius is related to $R_\Delta$ and $R_\gamma$ through the dimensionless parameter
\begin{equation}
\label{eq:xik}
    \xi_k \equiv \left(\frac{l_S}{\Delta_0}\right)^{1/2}\Gamma_0^{-\frac{4-k}{3-k}}~~,
\end{equation}
and in the case of $0\leq k<3$, we have \citep{yi2013}:
\begin{equation}
    \xi_k^{\frac{2}{4-k}}R_\Delta \sim R_\gamma \sim \xi_k^2 R_{\rm sp}~~.
\end{equation}
For the ``thick shell'' case, $\xi_k < 1$, the order of the radii is $R_\gamma < R_\Delta < R_\mathrm{sp}$, which means that the radial spreading of the shell is unimportant.
For the ``thin-shell'' case, $\xi_k>1$, the ordering is $R_\mathrm{sp}< R_\Delta < R_\gamma$, which means that the radial spreading is important \citep{sari1995}.
In this case, the shell spreads and reaches $\xi_k\sim 1$ before the reverse shock has crossed the shell.
As a result, the system soon satisfies $R_\Delta\sim R_\gamma$ and enters the deceleration phase that follows the BM solution \citep{blandfordmckee1976}.

Interestingly, for $0\leq k<2$ or $2<k<3$, the three characteristic timescales
$T_\mathrm{tr}$, $T_\mathrm{BM}$ and $T_\gamma$ for an on-axis observer are related as
\footnote{
If $k=2$, the transition phase has the same dynamical scaling as the coasting phase $\Gamma_\mathrm{const.}$, and therefore a distinct transition time $T_\mathrm{tr}$ cannot be defined.
In this case, equation (\ref{eq:Ttra}) is not applicable, and only the relation between $T_\gamma$ and $T_\mathrm{BM}$ remains meaningful.
}
\begin{equation}
\label{eq:timescales}
    \left(\frac{12-4k}{\xi_k}\right)^{\frac{1}{2-k}}T_\mathrm{tr} \sim T_\gamma \sim \frac{\xi_k^2}{2(C_\mathrm{RS}+\sqrt{3})}T_\mathrm{BM}~~.
\end{equation}
As examples that are relevant to our interests,
we here briefly give the following ordering of these timescales.
When $\xi_k < 12-4k$ and $0\leq k<2$, 
we have $T_\mathrm{tr}<T_\gamma$.
Furthermore, in the case of $\xi_k < 2(C_\mathrm{RS}+\sqrt{3})\sim4$, 
we obtain $T_\gamma < T_\mathrm{BM}$.
In particular, the second equality in Eq.~(\ref{eq:timescales}) shows that the smaller $\xi_k$ is, 
the more strongly the transition to the BM phase is delayed relative to $T_\gamma$.
Therefore, the boundary defined by $\xi_k = 1$, which characterizes the radial
dynamical behavior in the lab frame, does not generally coincide
with the threshold that determines the ordering of the characteristic observer time.
As a result, different regimes of $\xi_k$ can lead to qualitatively
distinct light-curve morphologies.

The observed synchrotron flux from the forward shock depends on each jet-dynamics phase: free expansion, transition and deceleration phases, which can be summarized as follows \citep{yi2013, gao2013}.
As in the standard manner, the peak frequency
$\nu_m$ and the cooling frequency $\nu_c$ measured by 
an on-axis observer are 
calculated  with microphysical parameters, including the fraction of shock internal energy transferred to electrons $\epsilon_e$, the fraction transferred to magnetic fields $\epsilon_B$, the power-law index of the electron energy distribution $p$, 
and the number fraction of electrons accelerated into the non-thermal distribution $f_\mathrm{e}$.

For the free-expansion dynamics described by Eq.~(\ref{eq:GammaFS_free}),
the theoretical flux density in the fast-cooling regime measured by an on-axis observer is given by
\begin{equation}
  F_{\nu, \mathrm{free, fast}}
  \propto
  \left\{
  \begin{array}{ll}\label{eq:free_fastcoolong_spec}
    t_{\rm obs}^{\frac{11-6k}{3}} & (\nu < \nu_c), \\
    t_{\rm obs}^{\frac{8-3k}{4}} & (\nu_c < \nu < \nu_m),\\
    t_{\rm obs}^{\frac{8-k(2+p)}{4}} & (\nu_m < \nu)~~,
  \end{array}
  \right.
\end{equation}
and in the slow-cooling regime, the flux scales as
\begin{equation}
  F_{\nu, \mathrm{free, slow}}
  \propto
  \left\{
  \begin{array}{ll}\label{eq:free_slowcoolong_spec}
    t_{\rm obs}^{\frac{9-4k}{3}} & (\nu < \nu_m), \\
    t_{\rm obs}^{\frac{12-(5+p)k}{4}} & (\nu_m < \nu < \nu_c),\\
    t_{\rm obs}^{\frac{8-(2+p)k}{4}} & (\nu_c < \nu).
  \end{array}
  \right.
\end{equation}
For the transition-phase dynamics described by Eq.~(\ref{eq:GammaFS_trans}),
the observed flux density  is
\begin{equation}
  F_{\nu, \mathrm{trans, fast}}
  \propto
  \left\{
  \begin{array}{ll}\label{eq:trans_fastcooling_spec}
    t_\mathrm{obs}^{\frac{16-9k}{3(4-k)}} & (\nu < \nu_c), \\
    t_\mathrm{obs}^{\frac{1}{2}} & (\nu_c < \nu < \nu_m),\\
    t_\mathrm{obs}^{-\frac{p-2}{2}} & (\nu_m < \nu)~~,
  \end{array}
  \right.
\end{equation}
in the fast-cooling regime, and
\begin{equation}
  F_{\nu, \mathrm{trans, slow}}
  \propto
  \left\{
  \begin{array}{ll}\label{eq:trans_slowcooling_spec}
    t_\mathrm{obs}^{\frac{16-7k}{3(4-k)}} & (\nu < \nu_m), \\
    t_\mathrm{obs}^{\frac{12-5k-(4-k)p}{2(4-k)}} & (\nu_m < \nu < \nu_c),\\
    t_\mathrm{obs}^{-\frac{p-2}{2}} & (\nu_c < \nu)~~,
  \end{array}
  \right.
\end{equation}
in the slow-cooling regime.
Finally, for the deceleration phase that follows the BM solution,
i.e., the dynamics described by Eq.~(\ref{eq:GammaFS_wind_1}),
the flux density in the fast-cooling regime is
\begin{equation}
  F_{\nu, \mathrm{bm, fast}}
  \propto
  \left\{
  \begin{array}{ll}\label{eq:BM_fastcoolong_spec}
    t_\mathrm{obs}^{\frac{2-3k}{3(4-k)}} & (\nu < \nu_c), \\
    t_\mathrm{obs}^{-\frac{1}{4}} & (\nu_c < \nu < \nu_m),\\
    t_\mathrm{obs}^{\frac{2-3p}{4}} & (\nu_m < \nu)~~,
  \end{array}
  \right.
\end{equation}
and in the slow-cooling regime, it becomes
\begin{equation}
  F_{\nu, \mathrm{bm, slow}}
  \propto
  \left\{
  \begin{array}{ll}\label{eq:BM_slowcoolong_spec}
    t_\mathrm{obs}^{\frac{2-k}{4-k}} & (\nu < \nu_m), \\
    t_\mathrm{obs}^{\frac{12-5k-3p(4-k)}{4(4-k)}} & (\nu_m < \nu < \nu_c),\\
    t_\mathrm{obs}^{\frac{2-3p}{4}} & (\nu_c < \nu)~~.
  \end{array}
  \right.
\end{equation}

In the thick-shell case ($\xi_k < 1$), 
it was found from Eq.~(\ref{eq:timescales}) that
$T_\mathrm{tr} < T_\gamma < T_\mathrm{BM}$ for $0<k<2$.
As an illustrative example, consider the following parameters:
$E_0=1\times10^{53}~\mathrm{erg}$, 
$\Gamma_0=100$, 
$\Delta_0=1\times10^{13}~\mathrm{cm}$,
$n_0=1~\mathrm{cm^{-2}}$, 
$k=1$, 
$p=2.3$, 
$\epsilon_e=1\times10^{-2}$, 
$\epsilon_B=1\times10^{-3}$, 
and $f_\mathrm{e}=1$.
Then, we have $\xi_k=0.66$,
$T_{\rm tr}=6.1$~s,
$T_{\gamma}=1.1\times 10^2$~s,
and
$T_{\rm BM}=1.3\times 10^3$~s,
so that
the optical light curve evolves as follows.
First, for $t_\mathrm{obs}<T_\mathrm{tr}$, the emission is in the slow-cooling regime,
and we get $F_\nu \propto t_\mathrm{obs}^{1.7}$.
Second, for $T_\mathrm{tr}<t_\mathrm{obs}<T_\mathrm{BM}$, the emission remains in the slow-cooling regime.
In this time range,
before minimum injection frequency $\nu_m$ crosses the optical band $\nu_\mathrm{opt}$, 
the light curve follows $F_\nu \propto t_\mathrm{obs}^{1.0}$,  
whereas after the crossing (at $t_\mathrm{obs}\approx60$~s) it becomes $F_\nu \propto t_\mathrm{obs}^{0.0}$.
Finally,
for $T_\mathrm{BM}<t_\mathrm{obs}$, 
the spectral regime initially satisfies $\nu_m<\nu_\mathrm{opt}<\nu_c$ and later transitions to $\nu_m<\nu_c<\nu_\mathrm{opt}$, leading to a temporal decay of
$F_\nu \propto t_\mathrm{obs}^{-1.1}$ and subsequently $F_\nu \propto t_\mathrm{obs}^{-1.2}$.
Like this example, in the transition phase $T_\mathrm{tr}<t_\mathrm{obs}<T_\mathrm{BM}$ with $k\sim1$, 
the observed light curve becomes nearly flat, and the breaks at $T_\mathrm{tr}$ and $T_\mathrm{BM}$ are achromatic.
We will show in Section~\ref{subsec:080330} that this characteristic 
 explains the observed optical/NIR afterglow features around the achromatic break of XRF~080330.

In the thin-shell case ($\xi_k \gg 1$), shell spreading becomes 
so important that $\xi_k$ approaches 
order unity \citep{yi2013, gao2013}, i.e.,
$T_\mathrm{tr}\sim T_\gamma\sim T_\mathrm{BM}$.
Around these times, the observed flux exhibits an achromatic peak.
We will show in Section~\ref{subsec:080710} 
that this fact accounts for the observed optical/NIR afterglow properties around the achromatic peak of GRB~080710.

Incorporation of 
the finite shell thickness allows 
the forward-shock dynamics to follow Eq.~(\ref{eq:GammaFS_trans_origin}), 
leading to 
the light curve behavior that can deviate from the analytical expectations in the thin shell limit.
This fact may provide a natural interpretation for observational 
results that are difficult to reconcile within the conventional thin-shell approximation.

\section{Numerical calculation and Bayesian inference}
\label{sec:codebaysian}

We compute
the afterglow emission using the open-source code \texttt{Magglow} \citep{kusafuka2025a, kusafuka202511}. 
This code calculates the synchrotron emission from the external shock based on the magnetic bullet afterglow model \citep{kusafuka2025a}.

The effects of the ejecta magnetization are neglected in this work.
In general, the magnetization of the ejecta is expected to mainly affect the acceleration phase of the outflow and the formation of the reverse shock \citep{kusafuka2025a, kusafuka202511}.
However, the afterglow data in this work do not show clear observational signatures that can be attributed to these effects.
Consequently, even if the initial magnetization parameter $\sigma_0$ were introduced as a free parameter, it would be difficult to constrain it meaningfully from the available observations.
For these reasons, the initial magnetization parameter is fixed to $\sigma_0=0$ throughout this study, and we do not discuss the detailed effects of ejecta magnetization.

In this calculation, the shell thickness is treated as a time-dependent quantity, whose evolution is calculated self-consistently based on the relative motion between the forward and reverse shocks. 
This treatment naturally incorporates the effects of shell spreading on the dynamical evolution and allows the thin-shell case ($\xi_k>1$) to be handled in a dynamically self-consistent manner.
In addition, the model includes the effects of the equal-arrival-time surface and Doppler boosting, enabling the relativistic line-of-sight effects for an arbitrary viewing angle, $\theta_\mathrm{obs}$ and a jet opening half-angle, $\theta_\mathrm{j}$, to be taken into account \citep{kusafuka202511}.

In order to place quantitative constraints on the 11 model parameters
($E_0$, $\Gamma_0$, $\Delta_0$, $n_0$, $k$, $\theta_\mathrm{j}$, $\theta_\mathrm{obs}$, $p$, $\epsilon_e$, $\epsilon_B$ and $f_\mathrm{e}$), 
we perform Bayesian inference in this study.
Bayesian inference provides a model-evaluation approach based on Bayes' theorem, 
which is expressed as 
\begin{equation}
    p(\vec{\Theta}|\vec{X}, H) = \frac{p(\vec{X}|\vec{\Theta})p(\vec{\Theta})}{p(\vec{X})}~~.
\end{equation}
Here, $\vec{X}$ denotes the observational data,
$\vec{\Theta}$ represents the parameter set of the model $H$,
$p(\vec{\Theta}|\vec{X}, H)$ is the posterior probability distribution,
$p(\vec{X}|\vec{\Theta}) \equiv {\cal L}(\vec{\Theta})$ is the likelihood,
$p(\vec{\Theta})$ is the prior distribution,
and $p(\vec{X}) \equiv {\cal Z}$ is the Bayesian evidence.
The Bayesian evidence is given by
\begin{equation}
  {\cal Z} = \int {\cal L}(\vec{\Theta})\, p(\vec{\Theta})\, d^D \vec{\Theta}~~,
\end{equation}
where $D$ is the number of model parameters.

To derive the posterior distributions, we employ the nested sampling algorithm \texttt{MultiNest} \citep{feroz2008, feroz2009, feroz2019}.
This method enables the computation of the Bayesian evidence ${\cal Z}$, allowing for a quantitative comparison between competing models.
Moreover, \texttt{MultiNest}  efficiently samples complex posterior distributions with multimodality and parameter correlations, which are commonly encountered in GRB afterglow models involving a large number of physical parameters.
As a result, the degeneracies and uncertainties inherent in the model can be directly quantified.
We utilize \texttt{PyMultiNest} \citep{pymultinest2016}, a Python interface to \texttt{MultiNest}.
In the sampling procedure, we use 500 live points and an evidence tolerance of 0.5  as the convergence criterion.

Assuming that the afterglow data follow Gaussian distributions, the likelihood function is defined as
\begin{equation}
    {\cal L}(\vec{\Theta}) 
  \equiv \prod_{s=1}^N \frac{1}{\sqrt{2\pi \sigma_s^2}} \exp \left({-\frac{(Y_s-F_s(\vec{\Theta}))^2}{2\sigma_s^2}}\right),
\end{equation}
where $N$ is the number of observational data points,
$Y_s$ and $\sigma_s$ are the median and standard deviation (1$\sigma$ error) of the $s$-th observed flux ($s=1,2, \cdots, N$), respectively, and
$\vec{\Theta}$ represents the model parameters, 
which are used to compute the theoretical flux $F_s$.
We adopt log-uniform priors for $E_0$, $\Gamma_0$, $\Delta_0$, $n_0$, $\epsilon_{e}$, $\epsilon_{B}$, 
and $f_{e}$, whereas we use uniform priors for $k$, $p$, $\theta_\mathrm{j}$ and $\beta=\theta_\mathrm{obs}/\theta_\mathrm{j}$.

\section{Results}
\label{sec:result}

In this section, we present the results of Bayesian inference for XRF~080330 and GRB~080710.
For the theoretical interpretation of the results, 
we adopt the maximum a posteriori (MAP) estimate,
defined as the parameter set that maximizes the posterior probability 
among the samples drawn during the Bayesian inference.
Analytical evaluations are performed based on this estimate.

\subsection{XRF 080330}\label{subsec:080330}

XRF~080330 was detected by \textit{Swift}, with a duration of 
$T_{90} = 67.0$~s in the 15--350~keV energy band \citep{guidorzi2009}. 
Its redshift is $z=1.51$ \citep{malesani2008}.
The peak energy in the cosmological rest frame is  
$(1+z)E_\mathrm{peak}< 88~\mathrm{keV}$, and
the isotropic-equivalent $\gamma$-ray energy is constrained as
$E_{\gamma, \mathrm{iso}}<2.2\times 10^{52}~\mathrm{erg}$ \citep{guidorzi2009}.
This event is classified as an XRF based on its fluence ratio, 
$S(25-50~\mathrm{keV})/S(50-100~\mathrm{keV})= 1.5^{+0.7}_{-0.3}$ \citep{guidorzi2009, sakamoto2008}.

In this study, we use optical/NIR afterglow data in the 
$g$-, $r$-, $i$-, $J$-, $H$- and $K$-bands, together with X-ray data at $3~\mathrm{keV}$.
All of these data are taken from \citet{guidorzi2009}.
The extinction toward XRF 080330 is known to be small both in the Milky Way and in the host galaxy, 
and is therefore neglected in our analysis \citep{guidorzi2009}.
The X-ray light curve initially exhibits a steep decay phase ($t<2\times10^2$~s), 
which is commonly attributed to the tail of the prompt emission and thus distinct from the 
external-shock afterglow emission \citep{kumar2000, tagliaferri2005, nousek2006, zhang2006, yamazaki2006}.

We performed Bayesian inference under several different conditions, 
varying the ranges of the prior distributions as well as the selection and relative weighting of the observational data,
in order to assess the robustness of the inferred parameters and to 
identify the most appropriate inference setup as our primary result.
Among these inference setups, several parameter constraints are found to be robust and largely insensitive to the adopted conditions.
First in \S~\ref{subsubsec:robustness}, 
we summarize the robust results and describe the rationale for selecting a specific inference setup as our primary result.
Next, a physical interpretation based on the corresponding 
MAP parameters in our primary result is presented
in \S~\ref{subsubsec:interpretation}.

\subsubsection{Common Results for Different Inference Setups and Selection of Primary Result}
\label{subsubsec:robustness}

Across all inference setups explored in this study, 
the inferred viewing geometry consistently satisfies 
$\beta \equiv \theta_{\rm obs}/\theta_\mathrm{j} < 1$.
This result implies a nearly on-axis configuration, 
with off-axis jet models being disfavored.
We will further discuss this in 
Section~\ref{subsec:noevidence_offaxis}.

One of the parameters that is particularly well constrained is the initial radial width, $\Delta_0$.
In all inference setups explored,
the posterior distribution of $\Delta_0$ remains confined to the range $8\times10^{12}~\mathrm{cm} \lesssim \Delta_0 \lesssim 1\times10^{13}~\mathrm{cm}$.
The tight constraint on $\Delta_0$ comes from
the observed achromatic break at $t_{\rm obs}\approx1.5\times10^3$~s 
in the optical/NIR light curves,
which corresponds to $T_{\rm BM}$ arising in the thick shell case.

Similarly, 
the slope of the CBM density profile, $k$, is also determined in the range,
$0.7 \lesssim k \lesssim 1.4$,
which is constrained by the temporal slopes of the light curve before the achromatic break.
We have performed Bayesian inference for models in which the density profile was fixed to $k=0$ (ISM model) or $k=2$ (wind model) using the same data set.
The resulting Bayesian evidence, $\ln {\cal Z}$, for the generalized density model with $k$ treated as a free parameter exceeds that of the ISM and wind models by more than $20$.
According to the Jeffreys' scale of evidence
\citep{Jeffreys1961}, 
this  difference in the Bayesian evidence indicates 
that this event strongly favors the generalized density model over the canonical ISM and wind scenarios.

\begin{figure*}[t]
	\centering 
  	\includegraphics[width=0.80\textwidth, angle=0]{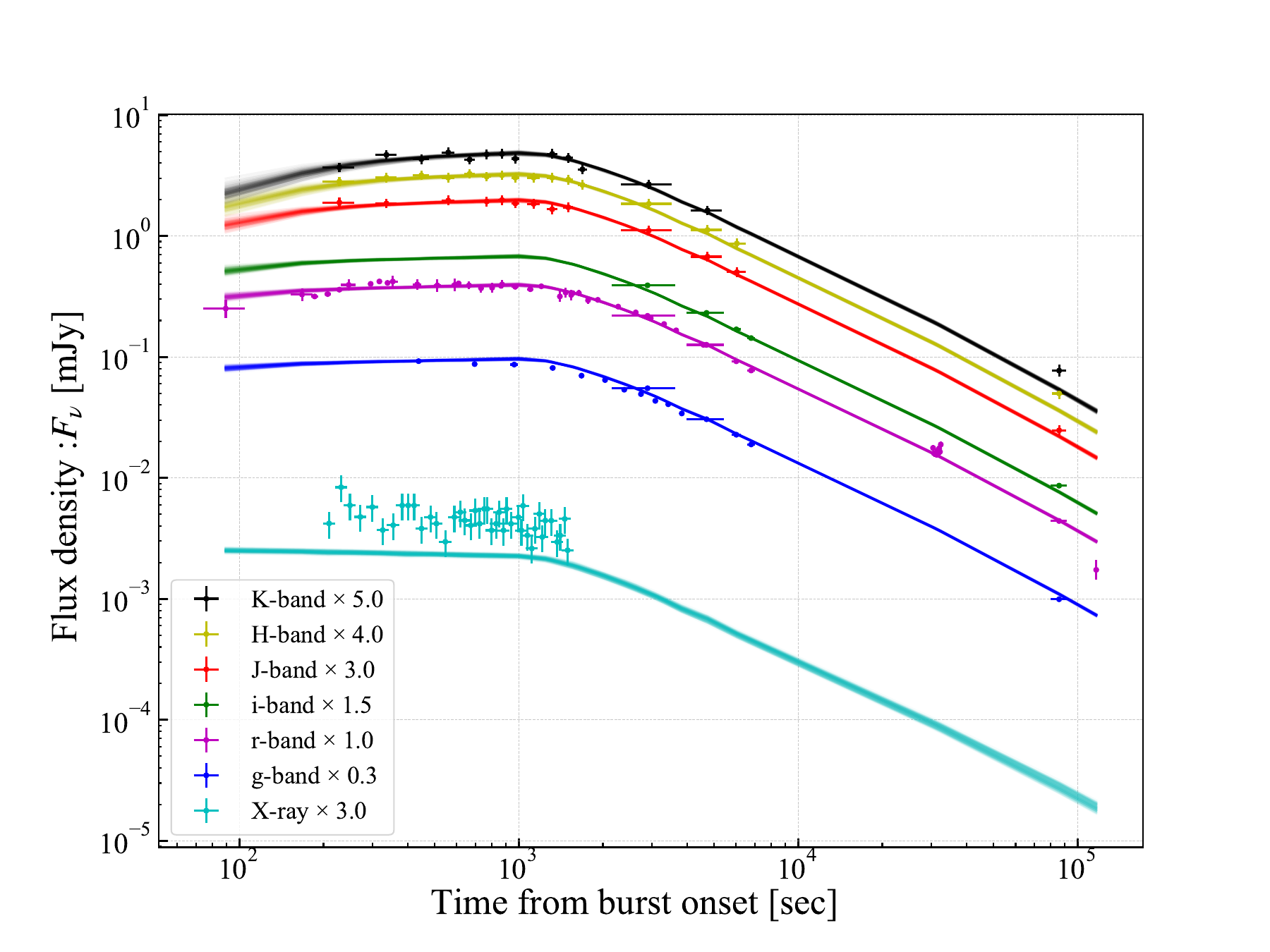}		
	\caption{
    Multi-wavelength afterglow observations of XRF 080330 (data points) together with 
    posterior predictive light curves for the primary result (100 faint solid lines).
    From top to bottom, the bands are the  
    $K$-band (black), $H$-band (yellow), $J$-band (red), $i$-band (green), $r$-band (purple), 
    $g$-band (blue) and the X-ray at 3 keV (cyan).
    The statistical weight of the X-ray data was reduced when deriving the preferred parameter set, while the original error bars are shown.
    } 
	\label{fig:080330_result_lc}%
\end{figure*}

In several of the inference setups explored, the forward-shock model successfully reproduces the optical/NIR light curves but systematically underpredicts the observed X-ray flux.
This discrepancy is expected because the present calculation assumes an infinitesimally thin emission region, 
which systematically underestimates the cooling frequency $\nu_c$ and hence suppresses the predicted X-ray flux (for more details, see \S~\ref{subsec:thinshell_emission}).
In addition, the observed X-ray emission may contain an extra component that is not directly associated with the forward-shock synchrotron emission responsible for the optical/NIR afterglow, such as residual internal emission or other high-energy processes \citep[e.g.,][]{ghisellini2007,yamazaki2009}.
These considerations suggest that the observed X-ray flux should be regarded as a weaker, upper-limit-like constraint on the forward-shock emission.
Motivated by these considerations, we conservatively assigned X-ray uncertainties twice as large as those reported by \citet{guidorzi2009} when deriving the preferred parameter set.
This treatment accounts for the expected systematic uncertainty in the X-ray flux prediction while allowing the fit to be primarily constrained by the optical/NIR data, which are well reproduced by the model\footnote{
Figure~\ref{fig:080330_result_lc} shows the original X-ray error bars reported by \citet{guidorzi2009}.
Only the statistical weights used in deriving the preferred parameter set were adjusted.
The reduced $\chi^2$ values quoted below were evaluated using the original X-ray uncertainties given by \citet{guidorzi2009}.
}.
With this approach, the inferred parameters become
$E_0\sim10^{54}$~erg and $\theta_\mathrm{j}\sim0.08$~rad
(see \S~\ref{subsubsec:interpretation}),
yielding physically plausible energetics while reproducing the observed optical/NIR light curves well.
For comparison, we also performed the inference using the original X-ray uncertainties.
In that case,
although the inferred shell width $\Delta_0$ remains  unchanged, 
the fit is driven toward extremely large values of
$E_0\sim10^{55}$~erg and $\theta_\mathrm{j}\gtrsim0.1$~rad,
resulting in an anomalously large collimation-corrected jet energy,
$E_{\rm K,jet}=E_0\theta_\mathrm{j}^2/2$,
which is energetically implausible
(see \S~\ref{sec:appendix1} for details).
We therefore adopt as our primary result the parameter estimates obtained with this conservative treatment, as they yield a more physically plausible interpretation while still adequately reproducing the multi-wavelength afterglow behavior.
The systematic effects of the thin emission region approximation on the inferred parameters and the X-ray flux are discussed further in Section~\ref{subsec:thinshell_emission}.

\begin{figure*}[t]
	\centering 
	\includegraphics[width=0.80\textwidth, angle=0]{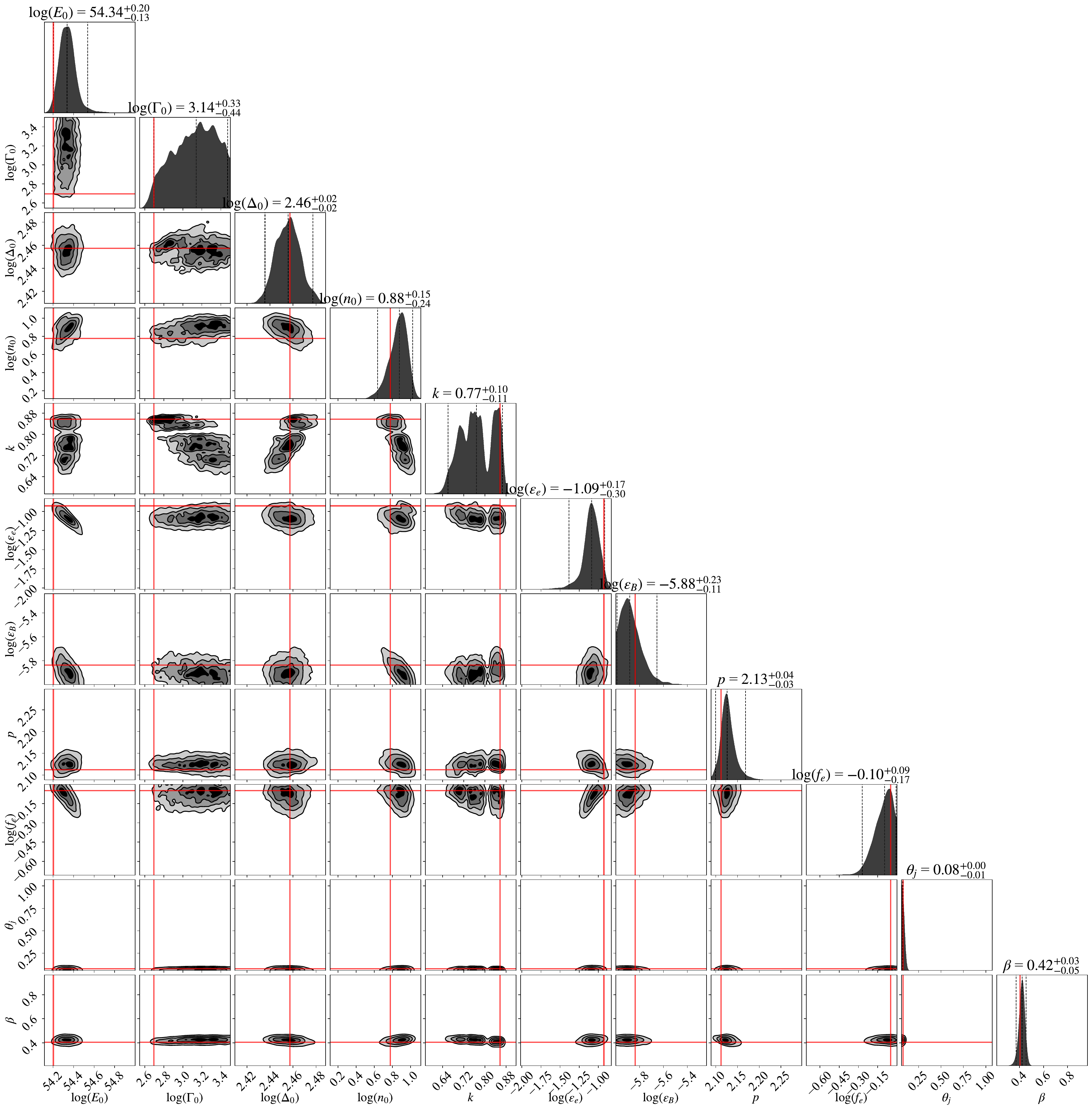}	
	\caption{
    Posterior probability distributions obtained from the Bayesian inference applied to the observational afterglow data of 
    XRF~080330 (primary result).
    The diagonal panels show the one-dimensional posterior distributions of each parameter,
    and the off-diagonal panels show the corresponding two-dimensional joint posterior distributions.
    The black dashed lines show the medians and the 95\% credible intervals, 
    and the red solid lines mark the maximum a posteriori (MAP) values;
    $E_0 = 1.6\times 10^{54}~\mathrm{erg}$, 
    $\Gamma_0=4.9\times 10^2$, 
    $\Delta_0= 8.6\times 10^{12}~\mathrm{cm}$, 
    $n_0 = 6.0~\mathrm{cm^{-2.14}}$,
    $k = 0.86$, 
    $\theta_\mathrm{j} = 0.08~\mathrm{rad}$,
    $\theta_\mathrm{obs} = 0.03~\mathrm{rad}$, 
    $p = 2.11$, 
    $\epsilon_e = 0.12$, 
    $\epsilon_B = 1.4\times 10^{-6}$, 
    and
    $f_\mathrm{e} = 0.89$.
    } 
	\label{fig:080330_result_corner}%
\end{figure*}

\subsubsection{Details of Primary Result}
\label{subsubsec:interpretation}

Table~\ref{table:result_XRF080330} summarizes the ranges of the prior distributions adopted to obtain our primary result, along with the medians and the 95\% credible intervals of the corresponding one-dimensional posterior distributions. 
The table also lists the 
MAP values:
$E_0 = 1.6\times 10^{54}~\mathrm{erg}$, 
$\Gamma_0=4.9\times 10^2$, 
$\Delta_0= 8.6\times 10^{12}~\mathrm{cm}$, 
$n_0 = 6.0~\mathrm{cm^{-2.14}}$,
$k = 0.86$, 
$\theta_\mathrm{j} = 0.08~\mathrm{rad}$,
$\theta_\mathrm{obs} = 0.03~\mathrm{rad}$, 
$p = 2.11$, 
$\epsilon_e = 0.12$, 
$\epsilon_B = 1.4\times 10^{-6}$, 
and
$f_\mathrm{e} = 0.89$.
Figure~\ref{fig:080330_result_lc} shows the observed afterglow data of XRF~080330, compared with 100 posterior predictive light curves (faint solid lines)
for parameter sets randomly sampled from the posterior distribution.
The model successfully reproduces the main observational features of the afterglow in the optical/NIR bands, including the gradual rise and the subsequent achromatic break.
The goodness of the fit evaluated at the MAP parameter set, 
including the X-ray data, is 
$\chi^2/\mathrm{d.o.f.}=669.6/(158-11)=4.5$ 
reflecting the known discrepancy between the X-ray and
optical/NIR data discussed above.
The corresponding value evaluated using only the optical/NIR data 
(i.e., excluding the X-ray data) is 
$\chi^2/\mathrm{d.o.f.}=367.6/(116-11)=3.5$.
Figure~\ref{fig:080330_result_corner} presents the corner plot showing the one- and two-dimensional posterior distributions corresponding to the primary result.

\begin{table}[t]
  \centering
  \caption{
  Model parameters, their prior ranges adopted in the analysis for primary result, 
  and the results of the Bayesian parameter estimation for XRF~080330.
  The marginalized one-dimensional posterior distributions are summarized by the median and the 95\% credible intervals, and the corresponding maximum a posteriori (MAP) values are given in the last column.
  }
  \renewcommand{\arraystretch}{1.22}
  \begin{tabular}{lccc}
    \hline
    Parameters & Prior Ranges & 1D dist & MAP\\
    \hline
    $\log(E_0~[\mathrm{erg}])$      & $[53, 55]    $ & $54.3^{+0.2}_{-0.1}$ & $54.2$\\
    $\log\Gamma_0$                & $[2, 3.5]      $ & $3.14^{+0.33}_{-0.44}$  & $2.69$ \\
    $\log(\Delta_0/c~[\mathrm{s}])$& $[2, 3]      $ & $2.46^{+0.02}_{-0.02}$  & $2.46$ \\
    $\log(n_0~[\mathrm{cm^{k-3}}])$ & $[-2, 2]     $ & $0.88^{+0.15}_{-0.24}$ & $0.78$ \\
    $k$                             & $[0, 2]      $ & $0.77^{+0.10}_{-0.11}$  & $0.86$ \\
    $p$                             & $[2.01, 2.5]$ & $2.13^{+0.04}_{-0.03}$  & $2.11$ \\
    $\log \epsilon_e$              & $[-3, -0.3]  $ & $-1.09^{+0.17}_{-0.30}$ & $-0.93$ \\
    $\log\epsilon_B$              & $[-6, -1]    $ & $-5.88^{+0.23}_{-0.11}$ & $-5.84$ \\
    $\log f_\mathrm{e}$                     & $[-1, 0]     $ & $-0.10^{+0.09}_{-0.17}$ & $-0.05$ \\
    $\theta_\mathrm{j}~[\mathrm{rad}]$       & $[0.001, 1.5]$ & $ 0.08^{+0.00}_{-0.01}$ & $0.08$ \\
    $\beta=\theta_{\mathrm{obs}}/\theta_\mathrm{j}$& $[0, 2]$ & $ 0.42^{+0.03}_{-0.05}$ & $0.40$ \\
    \hline
  \end{tabular}
  \label{table:result_XRF080330}
\end{table}

Using the MAP values, we obtain $\xi_k = 0.10$ 
so that 
this event satisfies the conditions for the thick-shell regime.
The onset time of the transition phase, 
given by Eq.~(\ref{eq:Ttra}), is calculated 
as $T_\mathrm{tr}\ll 1$~s, and
its end time is estimated from Eq.~(\ref{eq:TBM}) to be $T_\mathrm{BM}\approx 1.8\times 10^3$ s.
The rising and flat portions of the light curves up to 
$\approx 2\times 10^{3}$ s therefore correspond to the transition phase.
During this period, the condition 
$n_\mathrm{CBM}\gtrsim n_\mathrm{ej}/\Gamma_0^2$ 
holds, and the approximation given by Eq.~(\ref{eq:GammaFS_trans}) is valid. 

The cooling frequency $\nu_c$ crosses below the X-ray band at $T_c\approx 20$ s, 
the typical frequency $\nu_m$ crosses the $K$-band at $T_m\approx 2\times 10^2$ s.
Thus, the system remains in the slow-cooling regime throughout the transition phase.
Initially, the frequency ordering is 
$\nu_\mathrm{opt}<\nu_m<\nu_c$, 
and it evolves to 
$\nu_m < \nu_\mathrm{opt}<\nu_c$ 
by $t_\mathrm{obs}\approx 2\times 10^2$ s.
Accordingly, the chromatic break seen in the optical/NIR light curves at $t_\mathrm{obs}\sim10^{2}$~s (Fig.~\ref{fig:080330_result_lc}) marks the epoch 
at which the $\nu_m$ crosses the optical/NIR bands.
Observationally, the achromatic nature of the afterglow is maintained up to $t_\mathrm{obs}\sim 10^5$ s, which requires that the cooling frequency $\nu_c$ does not pass through the optical/NIR bands during this period.
Such behavior is realized if the value of  $\epsilon_B$ is 
small ($\sim10^{-6}$).

The jet opening half-angle and the viewing angle are constrained to 
$\theta_\mathrm{j} \approx 0.08~\mathrm{rad}$ and 
$\theta_\mathrm{obs}\approx 0.03~\mathrm{rad}$, respectively.
The constraints on $\theta_\mathrm{j}$ and $\theta_\mathrm{obs}$ are primarily driven by the late-time data around $t_\mathrm{obs}\sim10^{5}$~s, which provides a weak but non-negligible constraint.
The jet break time is given by \citep{rhoads1999, wei2003, decolle2012}
\begin{equation}\label{eq:Tjet}
    T_\mathrm{jet} \sim (1+z)|\theta_{\rm obs}\pm \theta_\mathrm{j}|^{\frac{8-2k}{3-k}}\frac{l_S}{2c}~~.
\end{equation}
Here, the plus (minus) sign corresponds to the epoch when emission from the jet edge farther from (closer to) the line of sight becomes visible to the observer.
Based on the MAP values, adopting the latter case corresponding to the jet edge closest to the line of sight, the jet break time is $T_\mathrm{jet}\approx7\times10^{4}$~s, 
consistent with the slightly steeper decay observed in the light curve after several $t_\mathrm{obs}\sim 10^{4}$~s.
The collimation-corrected kinetic energy of the jet is estimated to
be $E_\mathrm{K,jet} = E_0 \theta_\mathrm{j}^2/2 \approx 5.1\times10^{51}~\mathrm{erg}$,
which is somewhat larger than
those for the events with similar observed 
peak energy $(1+z)E_\mathrm{peak}<88~\mathrm{keV}$ 
of the prompt emission \citep{zhao2020}.

\subsection{GRB 080710}
\label{subsec:080710}
GRB~080710 was detected by \textit{Swift}, with a duration of $T_{90}=120$ s in the $15$--$350$ keV energy range, and is classified as a long GRB.
Its redshift is $z=0.845$ \citep{GCN-7962}.
The isotropic-equivalent gamma-ray energy and the rest-frame peak energy of the $\nu F_\nu$ spectrum were estimated as $E_{\gamma,\mathrm{iso}}=6\times10^{51}$ erg and $(1+z)E_{\rm peak}\sim2\times10^{2}$ keV, respectively \citep{kruhler2009b}.
The fluence ratio was $S(25-50~\mathrm{keV})/S(50-100~\mathrm{keV})=0.70\pm 0.15$ \citep{kruhler2009b}, so that
the burst was classified as a classical GRB, with errors ranging to a fluence ratio similar to those of 
X-ray rich GRBs when applying the definition of \citet{sakamoto2008}.

We use the same observational dataset as adopted in \citet{obayashi2024}:
$z$-band ($3.3\times 10^{14}$~Hz), 
$r$-band ($4.8\times 10^{14}$~Hz), 
and X-ray band ($0.3$~keV) \citep{kruhler2009b}.
Data thinning and the addition of systematic errors were also performed according to \citet{obayashi2024}.
The extinction toward GRB~080710 in $z$ and $r$-band is applied as 
$A_z = 0.128$~mag and $A_r = 0.220$~mag \citep{obayashi2024}.

We performed Bayesian inference under several different conditions by varying the prior distributions.
When all data points were treated with their original statistical uncertainties, the inference did not converge to a physically meaningful solution that reproduced the achromatic rise.
This is because the decaying phase contains many more data points than the rising phase, causing the likelihood function to be dominated by the late-time decay behavior.
Since the main purpose of the present analysis is to investigate the physical origin of the achromatic rise, 
we introduced an additional systematic uncertainty for the late-time data  used in the inference.
Specifically, for the $r$-band, $z$-band, and X-ray data 
in the decaying phase
($t_{\rm obs}>2.2\times 10^3$~s), we increased the uncertainties by a factor of ten.
This treatment accounts for the imbalance in the number of data points between the rising and decaying phases, and for the incompleteness of the adopted model.
In particular, the present model does not include jet structure or detailed late-time jet dynamics, both of which may affect the late-time evolution of the light curve.
Furthermore, the rising phase is the regime most directly relevant to distinguishing between the physical scenarios considered in this work.

Our present treatment with enlarged uncertainties in the decaying phase allows the inference to be primarily constrained by the rising part of the light curve, which is the feature most directly related to the transition from the energy-extraction phase to the 
Blandford--McKee phase\footnote{
Similar to the case of Fig.~\ref{fig:080330_result_lc},
we show in Fig.~\ref{fig:080710_result_lc} 
the observational data points with the 
same uncertainties as those of \citet{kruhler2009b}.
The reduced $\chi^2$ values quoted in the following 
were evaluated using the original observational uncertainties given by \citet{kruhler2009b}, rather than the enlarged uncertainties 
to get the primary result described 
in \S~\ref{subsubsec:interpretation2}.
}.
Using this treatment, the preferred parameter set achieves a lower reduced $\chi^2$ value than that obtained in \citet{obayashi2024}, indicating an improved agreement between the model and the observed light curves.
We refer to this case as the primary result.

\subsubsection{Details of Primary Result}
\label{subsubsec:interpretation2}

\begin{figure*}[t]
	\centering 	
	\includegraphics[width=0.80\textwidth, angle=0]{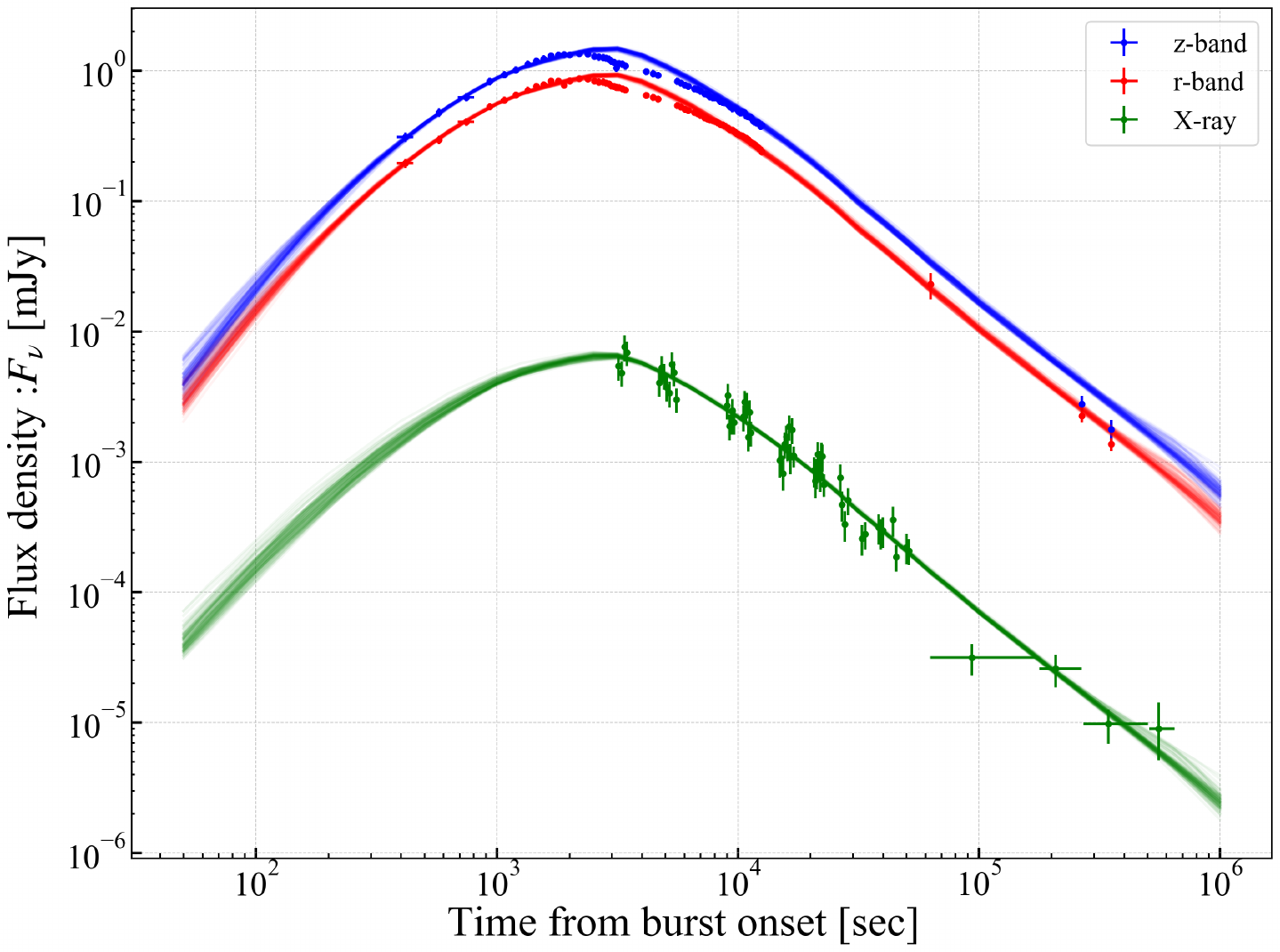}	
	\caption{
    Multi-wavelength afterglow observations of GRB 080710 (data points) together with 
    posterior predictive light curves (100 faint solid lines)
    for the primary result.
    From top to bottom, the bands are the  
    $z$-band (blue), $r$-band (red) and the X-ray at 0.3 keV (green).
    } 
	\label{fig:080710_result_lc}%
\end{figure*}

Table~\ref{table:result_GRB080710} summarizes the ranges of the prior distributions adopted in this analysis, along with the medians and the 95\% credible intervals of the corresponding one-dimensional posterior distributions, as well as the MAP values:
$E_0 = 2.1\times10^{54}~\mathrm{erg}$,
$\Gamma_0=42$,
$\Delta_0=1.3\times10^{13}~\mathrm{cm}$,
$n_0=1.1\times10^{2}~\mathrm{cm^{-2.95}}$,
$k=0.06$,
$\theta_\mathrm{j}=0.15~\mathrm{rad}$,
$\theta_{\rm obs}=0.01~\mathrm{rad}$,
$p=2.05$,
$\epsilon_e=2.2\times10^{-3}$,
$\epsilon_B=7.4\times10^{-2}$,
and $f_\mathrm{e}=0.85$.
Figure~\ref{fig:080710_result_lc} shows
the light curves of GRB~080710.
Within the range of model uncertainties corresponding to the posterior distributions, 
the model captures the gradual achromatic rise and the characteristic timescale of the peak.
The goodness of fit evaluated at the MAP parameter set for the rising phase is 
$\chi^2/\mathrm{d.o.f.}=43.3/(30-11)=2.2$,
and the $\chi^2/\mathrm{d.o.f.}$  evaluated using the full data set is 
$\chi^2/\mathrm{d.o.f.}=1002/(196-11)=5.4$.
Both values indicate an
improvement compared to the results obtained by our previous work \citep{obayashi2024}.
However, 
as shown in Fig.~\ref{fig:080710_result_lc},
the detailed behavior around the optical peak and in the subsequent decaying phase is still not fully reproduced.
This discrepancy suggests that jet structure effects may contribute to the observed flux during the late phase, indicating that the top-hat jet approximation adopted in this study may not be fully adequate for describing the late-time afterglow of GRB~080710.

\begin{table}[t]
  \centering
  \renewcommand{\arraystretch}{1.22}
  \caption{The same as Table~1, but for GRB 080710.}
  \begin{tabular}{lccc}
    \hline
    Parameters & Prior Ranges & 1D dist & MAP\\
    \hline
    $\log(E_0~[\mathrm{erg}])$      & $[50, 55]    $ & $54.1^{+0.5}_{-0.9}$ & $54.3$ \\
    $\log\Gamma_0$                & $[1, 3]      $ & $1.68^{+0.22}_{-0.11}$  & $1.62$ \\
    $\log(\Delta_0/c~[\mathrm{s}])$& $[1.5, 3.5]  $ & $2.67^{+0.05}_{-0.04}$  & $2.65$ \\
    $\log(n_0~[\mathrm{cm^{k-3}}])$ & $[-3, 3]     $ & $1.35^{+1.23}_{-1.66}$  & $2.05$ \\
    $k$                             & $[0,2]       $ & $0.03^{+0.08}_{-0.02}$  & $0.06$ \\
    $p$                             & $[2.01, 2.99]$ & $2.09^{+0.05}_{-0.03}$  & $2.05$ \\
    $\log\epsilon_e$              & $[-3,-0.3]   $ & $-2.55^{+1.00}_{-0.42}$ & $-2.65$ \\
    $\log\epsilon_B$              & $[-5,-0.3]   $ & $-1.37^{+0.89}_{-2.39}$ & $-1.13$ \\
    $\log f_\mathrm{e}$                     & $[-3, 0]     $ & $-0.49^{+0.46}_{-0.85}$ & $-0.07$ \\
    $\theta_\mathrm{j}~[\mathrm{rad}]$       & $[0.001, 1.5]$ & $0.21^{+0.18}_{-0.08}$  & $0.15$ \\
    $\beta=\theta_{\mathrm{obs}}/\theta_\mathrm{j}$& $[0,2]       $ & $0.77^{+0.09}_{-0.10}$  & $0.64$ \\
    \hline
  \end{tabular}
  \label{table:result_GRB080710}
\end{table}

\begin{figure*}[t]
	\centering 
	\includegraphics[width=0.80\textwidth, angle=0]{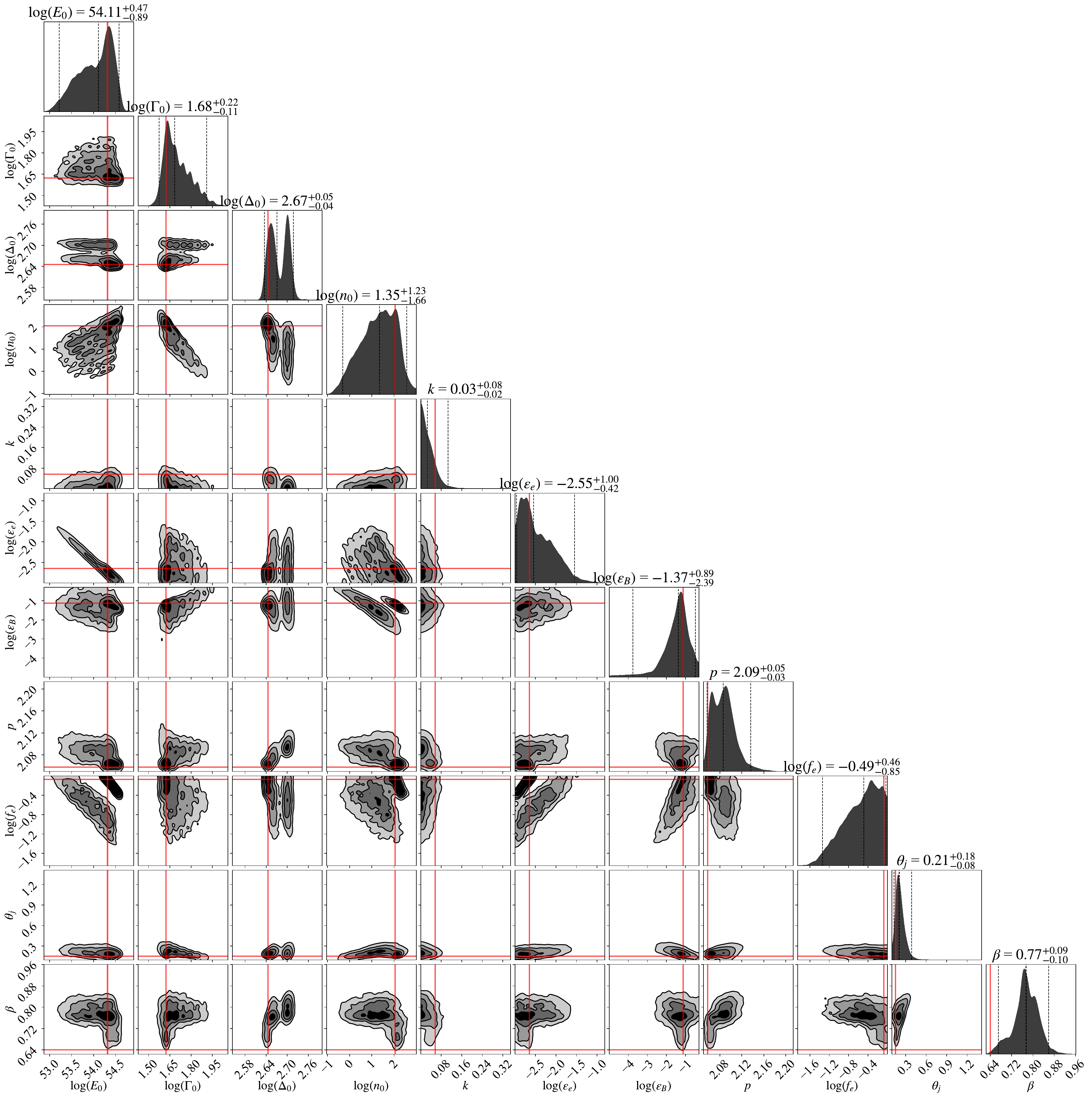}	
	\caption{
    Posterior probability distributions for GRB~080710, shown in the same format as Figure~\ref{fig:080330_result_corner}
    (primary result).
    The maximum a posteriori (MAP) values, shown as red solid lines, are
    $E_0 = 2.1\times10^{54}~\mathrm{erg}$,
    $\Gamma_0=42$,
    $\Delta_0=1.3\times10^{13}~\mathrm{cm}$,
    $n_0=1.1\times10^{2}~\mathrm{cm^{-2.95}}$,
    $k=0.06$,
    $\theta_\mathrm{j}=0.15~\mathrm{rad}$,
    $\theta_{\rm obs}=0.01~\mathrm{rad}$,
    $p=2.05$,
    $\epsilon_e=2.2\times10^{-3}$,
    $\epsilon_B=7.4\times10^{-2}$,
    and $f_\mathrm{e}=0.85$.
    } 
	\label{fig:080710_result_corner}%
\end{figure*}

The one-dimensional posterior distribution of $\log(\Delta_0/c)$ is bimodal, which can be seen in Figure~\ref{fig:080710_result_corner}. 
This bimodality appears only as a weak correlation
in the two-dimensional posterior distributions with $p$ and $\beta$.
Furthermore, the allowed range is relatively narrow, 
$\Delta_0/c=(4.3$--$5.2)\times 10^2$ s.
We therefore interpret this bimodality not as representing physically distinct solutions, but rather as an apparent effect arising from the geometry of the parameter space, such as marginalization and narrow allowed regions.
Substituting the obtained MAP values into Eq.~(\ref{eq:xik}), we find $\xi_k=2.2$, 
and the event is therefore classified as a thin-shell case ($\xi_k>1$) by definition, despite the initial radial width $\Delta_0$ being large.
This is because the shell classification is determined not only by the radial width $\Delta_0$ but also by the $\Gamma_0$. 
In this case, a small initial Lorentz factor increases the value of $\xi_k$, leading to $\xi_k>1$.
The preference for smaller $\Gamma_0$ is driven by the requirement to reproduce the observed peak time and the moderate rising slope of the optical light curve.
Therefore, the thin-shell classification in this event
should not be interpreted as implying a negligibly thin ejecta,
but rather as a consequence of the small Lorentz factor
required by the observed light-curve shape. 

To clarify how the rising phase is produced in this thin-shell case,
we examine the relevant timescales.
Using Eqs.~(\ref{eq:Ttra}) and (\ref{eq:TBM}), 
we estimate the beginning and the end of the transition phase as
$T_\mathrm{tr}\approx 1.3\times 10^3$ s and 
$T_\mathrm{BM}\approx 2.1\times 10^3$ s, respectively,
so that $T_\mathrm{tr}<T_\mathrm{BM}$.
Furthermore, we obtain $T_\gamma\approx 2.1\times 10^3$ s.
These three timescales are comparable,
suggesting that the observed rising part originates from emission produced during the gradual transition from the free-expansion phase to the transition phase.

From the MAP parameter values, the typical frequency already satisfies 
$\nu_m < \nu_\mathrm{opt}$ during the free-expansion phase.
The cooling frequency $\nu_c$  crosses below the $z$-band at $T_c\approx 50$ s.
Accordingly, after $t_\mathrm{obs}\approx 50$ s, 
the frequency ordering becomes $\nu_m<\nu_c<\nu_\mathrm{opt}<\nu_X$, 
under which the optical and X-ray light curves exhibit a steep rise close to $F_\nu\propto t_\mathrm{obs}^2$.
Subsequently, the ejecta are subject to weak deceleration up to the onset of the transition phase at $t_\mathrm{obs}\approx 10^3$ s.
During this period, the numerically calculated light curves show a more gradual rise, $F_\nu \propto t_\mathrm{obs}^1$, which is consistent with the observation result.
After a short transition phase, the dynamics approach the BM scaling at around $t_\mathrm{obs}\approx 2\times 10^3$ s, and the light curves enter a decaying phase.
The relation $\nu_m < \nu_c < \nu_\mathrm{opt}$ causes from relatively large values of 
$\epsilon_B$ and $f_\mathrm{e}$, and smaller values of $\epsilon_e$.
Indeed, this tendency is qualitatively consistent with that reported by \citet{obayashi2024}.

The posterior distribution of the viewing geometry converges to $\beta\equiv\theta_\mathrm{obs}/\theta_\mathrm{j}<1$,
indicating a nearly on-axis configuration and providing no support for off-axis afterglow.
Based on the MAP values, the jet break time [Eq.~(\ref{eq:Tjet})] is 
estimated as
$T_\mathrm{jet}\sim 10^6$~s for the jet edge farther from the line of sight and $T_\mathrm{jet}\sim 10^4$~s for the edge closer to the observer.
The latter is marginally consistent with the mild steepening observed in the late-time light curve.
The collimation-corrected kinetic energy of the jet is estimated to be $E_\mathrm{K, jet}=E_0\theta_\mathrm{j}^2/2\approx 2.8\times 10^{52}$ erg, which is slightly larger than that for the events with similar peak energy $(1+z)E_\mathrm{peak}=2\times 10^2$~keV 
of the prompt emission \citep{zhao2020}.

In this subsection, we have implicitly assumed for GRB~080710
that the observed X-ray afterglow and the optical/NIR 
afterglows originate from the same forward shock emission.
Then, we obtain a tight constraint 
on the electron energy spectral index, $p=2.09^{+0.05}_{-0.03}$.
The corresponding X-ray photon index is predicted to be $p/2+1=2.05^{+0.02}_{-0.02}$, which is consistent with the 
observed value of $1.92^{+0.12}_{-0.11}$. 
We note, however, that this constraint may be relaxed if the X-ray emission includes contributions from the other components than the forward-shock synchrotron emission or if the jet-structure effects become significant at late times.
Thus, GRB~080710 should be regarded as an illustrative case showing that finite-shell dynamics can improve the description of the achromatic rise.
At the same time, the present top-hat, forward-shock-only model should not be interpreted as a complete description of the entire afterglow light curve.
A more detailed treatment including jet-structure effects and possible additional emission components will be required to reproduce the detailed behavior around and after the optical peak.

\section{Discussion}
\label{sec:discussion}
In this paper, 
we have analyzed the multi-wavelength afterglow data of XRF 080330 and GRB 080710 using Bayesian inference, adopting an afterglow model that incorporates both a finite shell thickness and a power-law density profile of CBM.
We have shown that the model reproduces 
the observed gradual rise and achromatic peak in
the optical/NIR bands in both events.
Our analysis disfavors
the  off-axis afterglow ($\beta=\theta_{\rm obs}/\theta_\mathrm{j}>1$), 
which was claimed by 
\citet{guidorzi2009} and \citet{kruhler2009b}
for XRF~080330 and GRB~080710, respectively.
Instead, dynamical evolution
plays a crucial role in shaping the observed afterglow light curves.
For XRF 080330, the initial shell thickness and the external density slope are well constrained, allowing us to classify this event as a thick-shell case.
In contrast, for GRB 080710, a thin-shell dynamics may be applicable,
and the rising emission is shown to be produced by an intermediate dynamical regime between the free-expansion and transition phases.
In the following, based on the results for both events 
presented in \S~\ref{sec:result}, 
we discuss the implications for the origin of XRFs and GRBs and highlight the new insights provided by this study.

\subsection{Validity of the assumption of an infinitesimally thin emission region}
\label{subsec:thinshell_emission}

In this study, the jet dynamics are computed by explicitly accounting for the finite radial thickness of the ejecta, 
whereas the emission region is treated as infinitesimally thin in the radial direction.
In reality, the shocked region downstream of the shock front is expected to have a finite radial extent.

Both analytical and numerical studies have demonstrated that assuming an infinitesimally thin emission region leads to a systematic underestimation of the cooling frequency $\nu_c$ \citep{granot2002b, kusafuka_inprep}.
Since the synchrotron flux at frequencies above the cooling break scales as $F_\nu \propto F_{\nu,\max}\nu_c^{1/2}$, 
an underestimated $\nu_c$ directly results in an underestimation of the flux density in the high-frequency regime. 

\begin{figure}[t]
	\centering 
	\includegraphics[width=0.52\textwidth, angle=0]{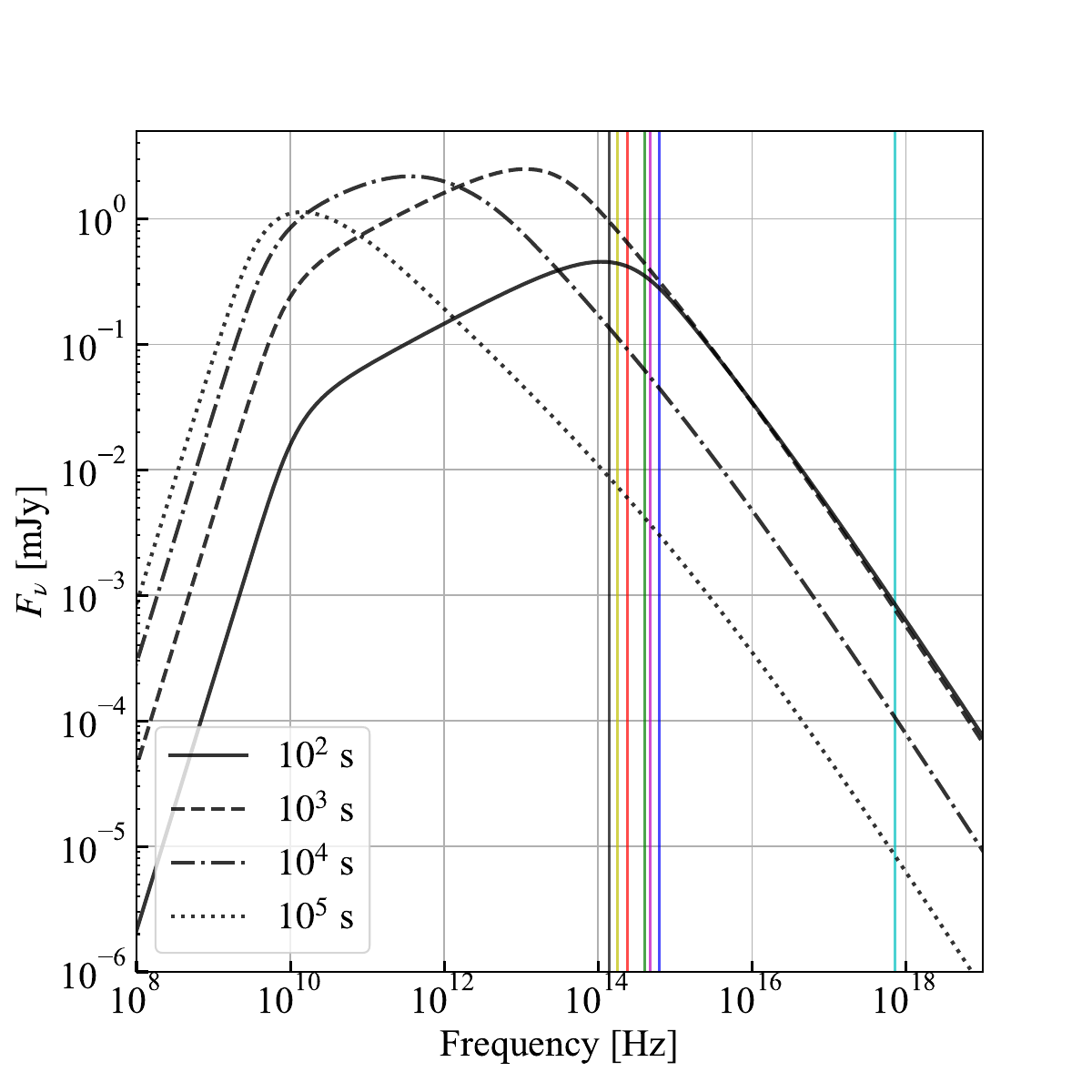}	
	\caption{
    SEDs with MAP parameter values of the primary result for XRF 080330 at observer time of 
    100 s (solid),
    $10^3$ s (dashed),
    $10^4$ s (dashed-dotted),
    and $10^5$ s (dotted) after the burst onset.
    The vertical solid lines indicate the frequencies corresponding to the X-ray band (3 keV) and the optical/near-infrared observing bands ($K$, $J$, $i$, $r$, and $g$). 
    The evolution of the spectral breaks can be seen from the temporal changes in the model spectra.
    } 
	\label{fig:080330_SED}%
\end{figure}

For XRF 080330, the X-ray band is located above the cooling frequency in our primary result.
Figure~\ref{fig:080330_SED} shows the spectral energy distributions (SEDs) corresponding to our primary result at four epochs. 
The cooling frequency is located at approximately 
$\nu_c\sim 10^{16}$~Hz, while the X-ray band ($\sim 10^{18}$~Hz) lies above the cooling break. 
In this regime, the model X-ray flux density may be systematically underestimated.
Indeed, as shown in Fig.~\ref{fig:080330_result_lc}, the theoretical light curve in the X-ray band is approximately a factor of two lower than the observed flux. 
Since the synchrotron flux above the cooling frequency scales as $F_\nu\propto \nu_c^{1/2}$, increasing $\nu_c$ by a factor of $\sim 4$ would increase the predicted X-ray flux by a factor of $\sim 2$.
This corresponds to the level of discrepancy between the model prediction and the observed X-ray flux shown in Fig.~\ref{fig:080330_result_lc}.
Therefore, the discrepancy may be alleviated by adopting a more accurate treatment of electron cooling that accounts for the finite radial extent of the emission region.

\begin{figure}[t]
	\centering 
	\includegraphics[width=0.52\textwidth, angle=0]{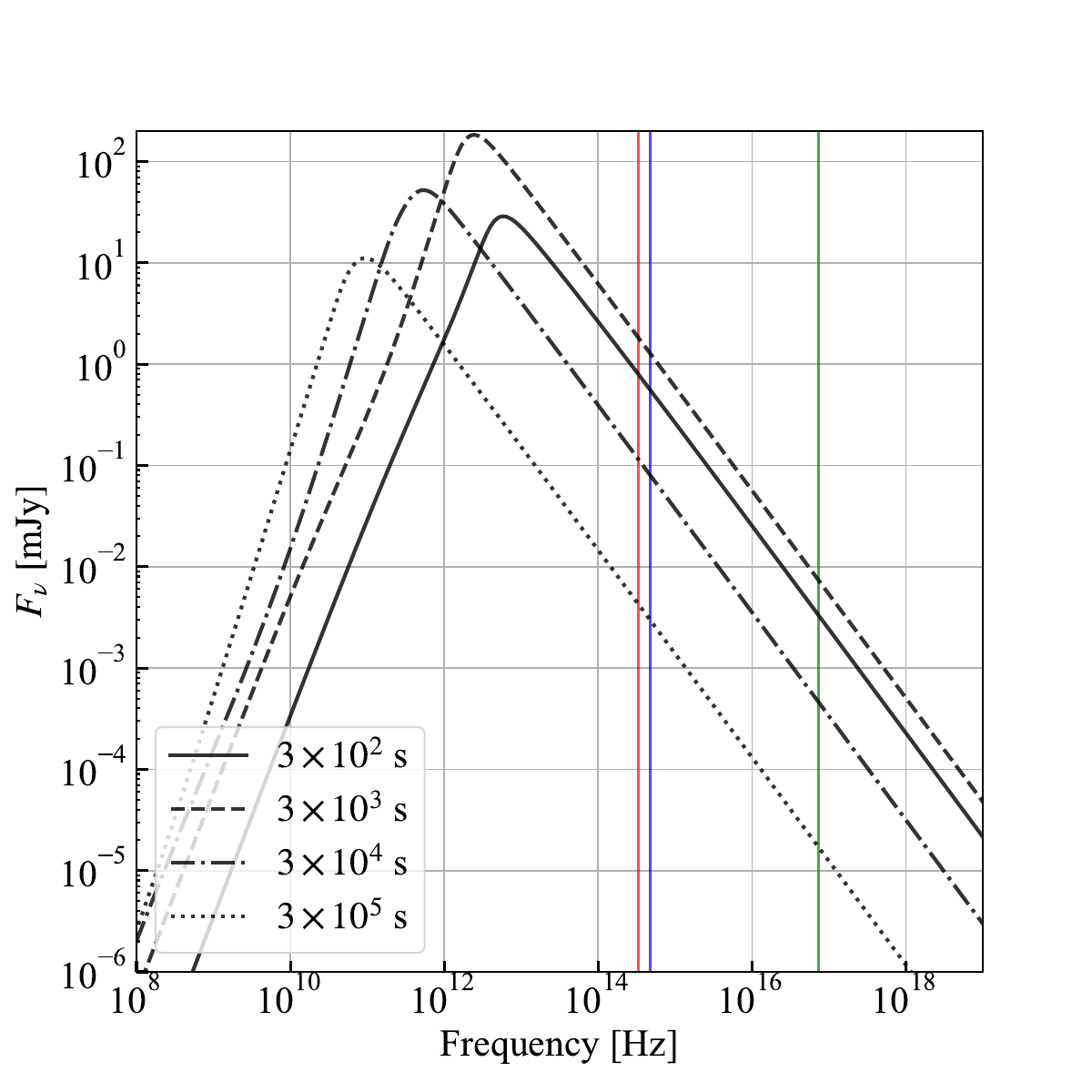}	
	\caption{
    SEDs for MAP parameter values of the primary result for GRB 080710 at observer time of 
    300 s (solid),
    $3\times 10^3$ s (dashed),
    $3\times 10^4$ s (dashed-dotted),
    and $3\times 10^5$ s (dotted) after the burst onset.
    The vertical solid lines indicate the frequencies corresponding to the X-ray band (0.3 keV) and the optical/near-infrared observing bands ($z$ and $r$). 
    } 
	\label{fig:080710_SED}%
\end{figure}

For GRB 080710, Figure~\ref{fig:080710_SED} shows the SEDs 
corresponding to our primary result at four epochs.
The characteristic frequencies satisfy
$\nu_m < \nu_c < \nu_\mathrm{opt} < \nu_X$
at observer times later than
$t_\mathrm{obs}\sim 10^2$ s.
In this case, both the optical/NIR and X-ray fluxes may be underestimated under the assumption of an infinitesimally thin emission region.
For $k=0$, during the free-expansion phase, the break frequency and normalization of the synchrotron spectrum scale as 
$\nu_c \propto \epsilon_B^{-3/2} n_0^{-3/2} \Gamma_0^{-4} t_{\rm obs}^{-2}$ and
$F_{\nu,\max} \propto f_\mathrm{e} \epsilon_B^{1/2} n_0^{3/2} \Gamma_0^{8}$, respectively.
In the deceleration phase, the corresponding scalings become
$\nu_c \propto \epsilon_B^{-3/2} n_0^{-1} E_0^{-1/2} t_{\rm obs}^{-1/2}$ and
$F_{\nu,\max} \propto \epsilon_B^{1/2} f_\mathrm{e} n_0^{1/2} E_0$
\citep{gao2013}.
Because $n_0$, $E_0$ and $\Gamma_0$ also affect the dynamical timescales, 
part of the systematic bias due to the assumption of an infinitesimally thin emission region may be mitigated by incorporating the finite radial extent of the shocked region in the radiation calculation.

Overall, 
if the cooling frequency is underestimated due to the assumption of an infinitesimally thin emission region, the inferred value of $\epsilon_B$ is likely underestimated, whereas $f_\mathrm{e}$ may be overestimated.
These systematic effects, however, do not qualitatively affect our primary conclusions regarding the constraints on the initial shell thickness $\Delta_0$ and the external density slope $k$.
A fully self-consistent treatment that incorporates a finite-thickness emission region in the parameter inference remains an important subject for future work.

\subsection{Disfavoring the off-axis jet scenarios for XRF~080330 and GRB~080710}
\label{subsec:noevidence_offaxis}
We find that the off-axis jet model, in which the observer is located outside the jet opening angle ($\beta=\theta_\mathrm{obs}/\theta_\mathrm{j} > 1$), is disfavored for both XRF~080330 and GRB~080710 based on Bayesian inference.
The posterior distributions show no support for the off-axis parameter region, 
in agreement with previous studies that tested off-axis structured jet models for GRB~080710 \citep{obayashi2024}.

The off-axis uniform jet model predicts several characteristic features in afterglow light curves arising from geometric effects 
\citep[e.g.][]{granot2002, totani2002, ramirez2005, granot2005, lamb2017, ryan2020}.
Specifically, the rising flux is expected to follow 
$F_\nu\propto t_\mathrm{obs}^{2-3}$ or even steeper, 
and an increasing viewing angle leads to a later peak time and a lower peak flux.
The observed afterglow light curves of XRF 080330 and GRB 080710 do not  match these expectations quantitatively.
%

\subsection{Active timescale of the central engine}
\label{subsec:central_engine}

Our analysis constrains the initial radial width of the ejecta to 
$\Delta_0\sim 10^{13}$~cm for both XRF~080330 and GRB~080710.
If the shell thickness reflects the duration of the central engine activity, 
the corresponding timescale is estimated to be
$\Delta_0/c\approx 300$~s for XRF~080330 
and $\approx 470$~s for GRB~080710 
(see Table~\ref{table:disscuss_CE_activity}).
These values are significantly longer than the 
duration of the prompt gamma-ray emission, $T_{90}/(1+z)\approx 24$~s 
and $\approx 65$~s for XRF~080330 and GRB~080710, respectively.
This suggests that the central engine activity inferred from 
the shell thickness is longer by a factor of $\sim 10$
than that implied by the prompt gamma-ray emission.
Such extended central engine activity timescales are commonly observed in GRBs.
Indeed, it has long been known that the duration of the prompt emission
in the X-ray band is substantially longer than that in the gamma-ray band \citep{murakami1991, jiang2025, guidorzi2025, intzand2025}. 
Recently, the Einstein Probe detected XRF~240801B (EP240801a), 
which exhibited a similarly prolonged X-ray duration \citep{jiang2025}.
Furthermore, X-ray flares detected in the X-ray afterglow are naturally interpreted as the result of a long-acting central engine.
\citep[e.g.][]{burrows2005, ioka2005, zhang2006, bernardini2011, margutti2010}.
These observations indicate that the central engine can remain active beyond the timescale 
defined by the prompt gamma-ray emission, placing our results within a broader observational context.

\begin{table}[t]
  \centering
  \caption{Comparison of estimated central engine activity times in rest frame based on prompt gamma duration and shell thickness.
   (References: (1) \citet{guidorzi2009} (2) \citet{kruhler2009b})
  }
  \label{table:disscuss_CE_activity}
  \renewcommand{\arraystretch}{1.22}
  \begin{tabular}{c|c|c}
    \hline
                    & XRF 080330    & GRB 080710 \\
    \hline
    $T_{90}/(1+z)$  & $24$~s~$^{(1)}$        & $65$~s~$^{(2)}$\\
    \hline
    $\Delta_0/c$    & $300$~s       & $470$~s\\
    \hline
  \end{tabular}
\end{table}

Our finding of $cT_{90}/(1+z)< \Delta_0$
for XRF~080330 and GRB~080710 suggests 
that the radial width of
the region of the prompt gamma-ray emission is smaller than 
the whole ejecta thickness,
indicating that the radial jet structure is not uniform.
For example, suppose that multiple shells with individual radial widths 
 $\Delta R\sim R/\Gamma^2$ are contained within the ejecta with
 the full width $\Delta_0$.
If only a fraction of these shells predominantly emit $\gamma$-rays while the remaining shells primarily produce X-ray emission, the effective radial extent of the $\gamma$-ray emitting region would be smaller than that of the entire ejecta. 
In this case, the duration of the prompt gamma-ray emission
would naturally be shorter than that in the X-ray band, 
and time differences of their arrival at an observer could arise 
from the relative locations of the emitting shells.

\subsection{Efficiency of the prompt emission}

Using the MAP values of the isotropic-equivalent kinetic energy $E_0$, 
we estimate the  efficiency of  the prompt gamma-ray emission 
in the standard manner as
$\eta_\gamma = E_{\gamma,\mathrm{iso}} /
(E_{\gamma,\mathrm{iso}} + E_0)$, and
we obtain $\eta_\gamma \leq 2.6\times10^{-2}$ for XRF~080330 and
$\eta_\gamma = 2.9\times10^{-3}$ for GRB~080710.
However, 
these estimates assume that the entire ejecta contribute to the prompt emission.
As discussed in Section~\ref{subsec:central_engine}, if only a fraction $f_\mathrm{rad}\sim cT_{90}/\Delta_0$ of the ejecta is involved in the prompt phase, the effective kinetic energy participating in the prompt emission should be reduced to $f_{\rm rad}E_0$.
In this case, 
the  ``local'' prompt radiative efficiency becomes
$\eta_{\gamma,\mathrm{local}} = 
E_{\gamma,\mathrm{iso}} /
(E_{\gamma,\mathrm{iso}} + f_{\rm rad}E_0)
\leq 0.25$ for XRF~080330 and
$\eta_{\gamma,\mathrm{local}} \simeq 0.02$ for GRB~080710.
This correction alleviates the extremely low efficiency implied by the
traditional estimate.

At the characteristic prompt emission radius 
$R\sim 10^{13}$--$10^{14}$~cm, the MAP values derived from our afterglow analysis imply the bulk Lorentz factor of 
$\Gamma\approx 3\times 10^2$ for XRF~080330 and 
$\Gamma\approx 4\times 10^1$ for GRB~080710.
The relatively large Lorentz factor inferred for XRF~080330 suggests 
internal shock emission from
low baryon-loading outflows with small velocity contrasts \citep{daigne2003}.
In such a scenario, 
the relative Lorentz factors  between colliding shells are expected to be small, 
leading to inefficient energy dissipation through internal shocks.
On the other hand, the lower Lorentz factor inferred for 
GRB~080710 points to an external shock emission of 
a dirty fireball regime \citep{dermer1999}.
In this case, the efficiency of the prompt gamma-ray emission
is much smaller than usual \citep{sari1997}.

\subsection{Possible CBM diversity as a consequence of different progenitor evolution}

Our analysis reveals a clear difference in the external density profiles surrounding XRF 080330 and GRB 080710.
We find that the external density distribution follows $n(R)\propto R^{-k}$ with $k\sim 1$ for XRF 080330, whereas GRB 080710 is consistent with an approximately uniform medium ($k\approx 0$).
This difference may indicate diversity in the circumstellar environments through which the GRB jets propagate.
We compare our results with the analytical study of \citet{yi2013}, 
who analyzed early afterglow light curves of 19 GRBs, 
including XRF 080330 and GRB 080710, assuming thin-shell and power-law density profiles.
They reported $k=1.32$ for XRF 080330 and $k=0.92$ for GRB 080710.
Their analysis relied on analytical approximations and qualitative comparisons with observed temporal and spectral indices.
Our result for XRF~080330 is roughly consistent with that of \citet{yi2013}, 
whereas our result for GRB~080710 is not consistent with theirs.

The CBM density profile, $n(R)\propto R^{-k}$,
depends on the time evolution of the mass-loss rate, 
$\dot{M}(t)$, and the wind velocity, $v_w(t)$,
of the progenitor.
If the wind velocity remains approximately constant, the value 
$k\sim 1$ for XRF~080330 implies that the mass-loss rate decreased toward the final stages prior to collapse.
If the mass-loss rate were roughly constant, the same result would require an increase in wind velocity toward core collapse.
Either scenario points to time-variable mass-loss behavior in the late evolution of the progenitor of XRF~080330.
In contrast, the  nearly uniform CBM inferred for GRB~080710 suggests that a wind-shaped density gradient was either not sufficiently established or was erased by prior mass ejection episodes, leading to an ISM-like environment \citep{vanMarle2006, schulze2011, chrimes2022}.
This difference between the two events likely reflects variations 
in the evolutionary stage of the progenitors immediately before collapse 
and in their interaction histories with the surrounding medium.
The density slope $k$ thus emerges as an important observational probe of the terminal evolution of GRB progenitors.

Using the inferred values of $n_0$ and $k$, 
we estimate the effective wind parameter $\dot{M}/v_w$ that shapes the environment at a characteristic radius $R\sim 10^{17}$ cm, 
where the afterglow emission is produced: 
\begin{equation}
    \frac{\dot{M}}{v_w}= 4\pi m_\mathrm{p} A(n_0, k)
    \left(\frac{R}{10^{17}\,\mathrm{cm}}\right)^{2-k}.
\end{equation}
Adopting the MAP values, we obtain 
$\dot{M}/v_w \approx 1.4\times10^{-8}$
$\mathrm{M_\odot\,yr^{-1}\,km^{-1}\,s}$ for XRF~080330 and
$\dot{M}/v_w\approx 7.9\times10^{-9}$
$\mathrm{M_\odot\,yr^{-1}\,km^{-1}\,s}$ for GRB~080710.
These values are consistent with the typical ranges inferred for progenitors of Type~II and stripped-envelope supernovae from radio observations \citep[e.g.,][]{sfaradi2025}.

\subsection{Radio afterglow}

Using the MAP values derived from our Bayesian analysis, we estimate that the $1.4$~GHz radio afterglow of XRF~080330 
peaks at $\sim 0.1$~mJy at $t \sim 1\times10^{5}$~s after the burst onset.
For GRB~080710, the predicted $1.4$~GHz emission peaks later, at $t \sim 6\times10^{5}$~s, with a much lower flux of $\sim 10^{-3}$~mJy.
No radio afterglow detections have been reported for either event \citep{guidorzi2009, kruhler2009b, chandra2012}.

For comparison, typical GRB radio afterglows at 1.4~GHz peak at 
$t \sim 10^{5}$--$10^{7}$~s with flux densities of $0.1$--$1$~mJy \citep{chandra2012}, while the $1.4$~GHz Expanded Very Large Array (EVLA) limiting sensitivity for a 30-minute integration is $\sim 0.05$~mJy \citep{chandra2012}.
Therefore, the radio afterglow of XRF~080330 might have been marginally detectable with facilities available at that time, 
whereas that of GRB~080710 might be too faint to be detected.

This contrast primarily arises from differences in the external density and microphysical parameters inferred for the two events.
In particular, the magnetic-field energy fraction $\epsilon_B$ differs by nearly four orders of magnitude, being significantly larger for GRB~080710 than for XRF~080330.
A stronger magnetic field required for GRB~080710 increases the synchrotron self-absorption frequency $\nu_a$ ($\sim10^{12}$~Hz), 
leading to substantial suppression of 1.4~GHz radio emission.
On the other hand, for
XRF~080330, we obtain $\nu_a\sim10$~GHz, for
MAP parameters.
Both events are in the slow-cooling regime, but the ordering of the characteristic frequencies differs: while XRF~080330 is consistent with $\nu_a < \nu_m < \nu_c$ until $\sim 10^5$ s, GRB~080710 likely satisfies $\nu_m < \nu_a < \nu_c$ from $\sim 10^2$ s to $\sim 10^7$ s, in which case emission in the 1--100~GHz range is strongly self-absorbed and therefore intrinsically faint.

These results indicate that the presence or absence of radio afterglow emission provides a sensitive probe of the circumburst density and microphysical parameters.
Future radio follow-up observations will be crucial for breaking degeneracies that remain when only optical and X-ray data are available, and will play an important role in testing and refining physical models of GRB afterglows.

\subsection{Constraints on the reverse-shock emission}
In the Bayesian inference performed in this work, we did not include reverse-shock emission for either XRF~080330 or GRB~080710. 
This is motivated observationally by the absence of a rapidly decaying phase in the optical/NIR bands, 
which is a characteristic signature of reverse-shock emission. Therefore, there is no direct observational motivation to introduce an additional reverse-shock component in the parameter estimation.
Nevertheless, for consistency, it is necessary to examine whether the jet and circumburst parameters inferred from the forward-shock-only analysis would predict a reverse-shock emission component that is brighter than the observed optical/NIR afterglow. In particular, if the reverse shock were sufficiently bright to exceed the observed flux, the forward-shock-only interpretation would not be self-consistent.

We evaluated the reverse-shock emission using the \texttt{Magglow} code.
The microphysical parameters of the reverse-shock region, $\epsilon_{e,\rm RS}$, $\epsilon_{B,\rm RS}$, $f_{e,\rm RS}$, and $p_{\rm RS}$, need not be identical to those of the forward shock. We therefore treat them separately from the forward-shock microphysical parameters. In the following consistency check, we fix the dynamical and geometrical parameters, $E_0$, $\Gamma_0$, $\Delta_0$, $k$, $n_0$, $\theta_{\rm j}$, and $\beta=\theta_{\rm obs}/\theta_{\rm j}$, to the MAP values obtained from the forward-shock-only inference. 
We also fix $p_{\rm RS}=2.5$ and $f_{e,\rm RS}=1$, and then determine the range of $\epsilon_{e,\rm RS}$ and $\epsilon_{B,\rm RS}$ for which the reverse-shock emission does not exceed the observed optical/NIR flux at any observed epoch.

For XRF 080330, we find that the reverse-shock emission remains below the observed optical/NIR flux for $\epsilon_{e,\rm RS}\lesssim0.06$, while $\epsilon_{B,\rm RS}$ is unconstrained.
For GRB 080710, the reverse-shock emission remains dimmer than
 the observed flux  in the case of $\epsilon_{e,\rm RS}\lesssim0.03$ and
$\epsilon_{B,\rm RS}\lesssim3\times10^{-5}$.
These results indicate that a reverse-shock component can be present within the allowed parameter range. 
However, the observed optical/NIR data constrain its contribution and disfavor a reverse shock sufficiently strong to dominate the observed afterglow light curves. 
Therefore, the inclusion of a reverse-shock component does not qualitatively alter our main conclusions regarding the shell thickness and the external density profile.

\section{Conclusion}

In this paper, 
we investigated the physical origin of the achromatic peaks observed several thousand seconds after the burst 
in the multi-wavelength afterglows of XRF~080330 and GRB~080710.
We performed model comparison and parameter estimation for XRF~080330 and GRB~080710 using the semi-analytic afterglow code \texttt{Magglow}, 
which incorporates both finite ejecta thickness and a power-law external density profile, and Bayesian inference based on the nested sampling algorithm implemented in \texttt{MultiNest}.

We find that for both events, the achromatic peak observed at several thousand seconds is best explained by jet dynamics rather than by off-axis viewing effects, implying a shell thickness of $\Delta_0\sim10^{13}$~cm for both bursts.
This scale is approximately an order of magnitude larger than the spatial scale inferred from $cT_{90}$, suggesting that the central engine activity lasts significantly longer than implied by the prompt gamma-ray emission alone.
Furthermore, Bayesian evidence favors a generalized external density profile over the canonical $k=0$ (ISM) and $k=2$ (steady wind) models.
While XRF~080330 is best described by a non-uniform medium with $k\sim 1$, 
GRB~080710 prefers an approximately uniform density profile with $k\sim0$, indicating diversity in the circumstellar environments shaped by the terminal evolution of the progenitor stars.

Overall, our results demonstrate that afterglow models that account for the finite thickness of the ejecta may be important for interpreting the temporal structure of early GRB afterglows and may provide new insights into the radial structure of GRB jets and the duration of central engine activity.
Since such constraints cannot be inferred from the prompt emission duration alone, this framework offers a crucial link between the prompt and afterglow phases.
Applying this approach to a larger sample of GRBs will enable a statistical assessment of both the universality and diversity of early afterglow emission mechanisms, emphasizing the importance of incorporating early afterglow dynamics in understanding the physical nature of GRB engines.

Finally, we
note that our parameter estimations rely on the assumption 
that the observed multi-wavelength afterglow emission is entirely produced by synchrotron radiation from the forward shock.
Future reanalyses that incorporate possible contributions from 
reverse-shock emission or structured jet models will be 
important for assessing the robustness of the inferred 
parameter values.

\section*{Acknowledgements}

We thank 
Makoto~Uemura,
Yohei~Sudo,
Hatsune~Goto,
Daisuke~Yonetoku,
Motoko~Serino,
Takanori~Sakamoto,
Kohta~Murase, 
and 
Shuta~J.~Tanaka 
for helpful comments and discussions. 
 We also thank the anonymous referee for his/her/their helpful comments
that improved the paper. 
This work was supported by JST SPRING, Grant Number JPMJSP2103
(KO).
This research was also supported by the joint research program of the Institute for Cosmic Ray Research (ICRR), the University of Tokyo,
and was partially supported by JSPS KAKENHI 
Grant Nos.~23K22522, 23K25907 (RY), 23H04899 (RY and KA),
JP23KJ0692 (YK), 24H00025, 25K07352 (KA).
Finally, the authors thank the Yukawa Institute for Theoretical Physics at Kyoto University. Discussions during the YITP long-term workshop YITP-T-26-02 on ``Multi-Messenger Astrophysics in the Dynamic Universe'' were useful to complete this work.

\appendix
\section{Parameter estimation for XRF 080330 using the original observational uncertainties}
\label{sec:appendix1}

\begin{figure*}[t]
	\centering 	
	\includegraphics[width=0.80\textwidth, angle=0]{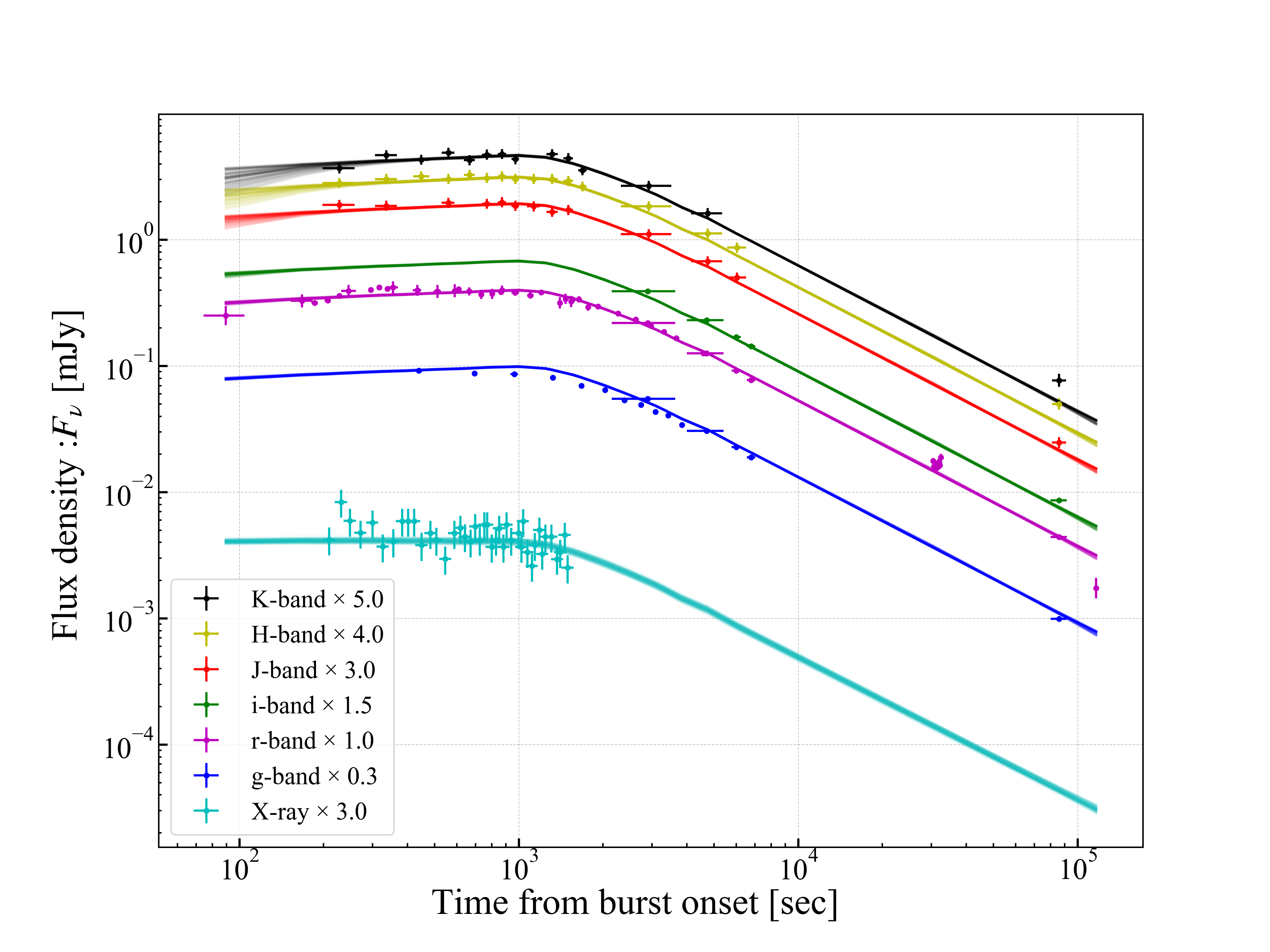}	
	\caption{
    Multi-wavelength afterglow observations of XRF 080330 (data points) together with 
    posterior predictive light curves (100 faint solid lines).
    This is the result for parameter inference without assigning larger uncertainties to the X-ray data.
    From top to bottom, the bands are the  
    $K$-band (black), $H$-band (yellow), $J$-band (red), $i$-band (green), $r$-band (purple), 
    $g$-band (blue) and the X-ray at 3 keV (cyan).
    } 
	\label{fig:080330_result_lc_X}%
\end{figure*}

\begin{figure*}[t]
	\centering 
	\includegraphics[width=0.80\textwidth, angle=0]{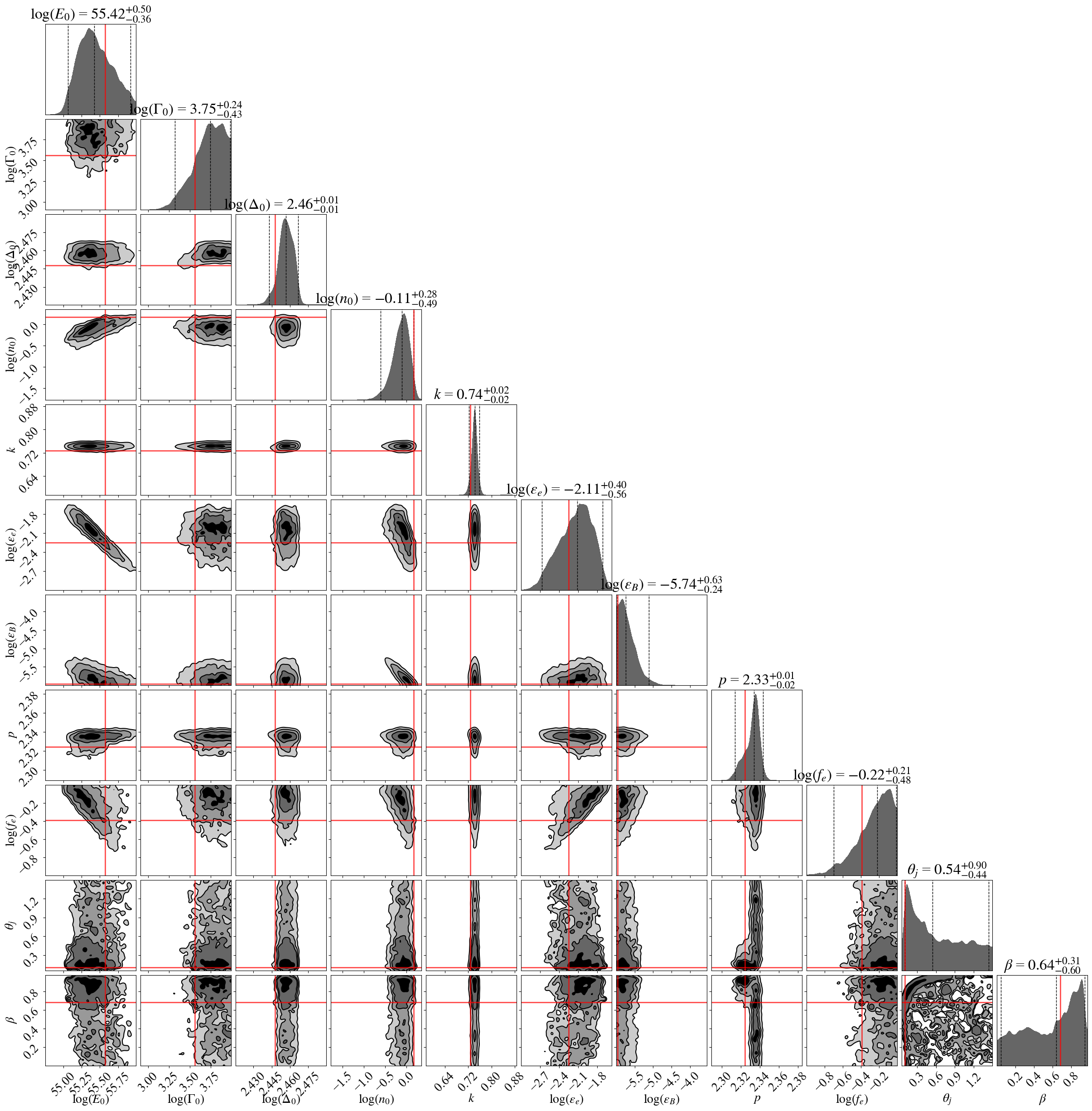}	
	\caption{
    Posterior probability distributions for XRF~080330, shown in the same format as Figure~\ref{fig:080330_result_corner}.
    This is the result for parameter inference without assigning larger uncertainties to the X-ray data.
    The maximum a posteriori (MAP) values, shown as red solid lines, are
    $E_0 = 3.7\times10^{55}~\mathrm{erg}$,
    $\Gamma_0=3.6\times 10^3$, 
    $\Delta_0= 8.4\times 10^{12}~\mathrm{cm}$, 
    $n_0 = 1.5~\mathrm{cm^{-2.27}}$,
    $k = 0.73$, 
    $\theta_\mathrm{j} = 0.10~\mathrm{rad}$,
    $\theta_\mathrm{obs} = 0.07~\mathrm{rad}$, 
    $p = 2.32$, 
    $\epsilon_e = 5.6\times10^{-3}$, 
    $\epsilon_B = 1.1\times 10^{-6}$, 
    and
    $f_\mathrm{e} = 0.41$.
    } 
	\label{fig:080330_result_corner_X}%
\end{figure*}

For completeness, we also performed parameter estimation for XRF 080330 using the original observational uncertainties reported by 
\citet{guidorzi2009},
without assigning larger uncertainties to the X-ray data. 
The resulting posterior distributions and the primary model are shown in 
Table~\ref{table:result_XRF080330_original}, 
and Figures~\ref{fig:080330_result_lc_X} and \ref{fig:080330_result_corner_X}.
The goodness of the fit evaluated at the MAP parameter set, including the X-ray data, is $\chi^2/{\rm d.o.f}=525.1/(158-11)=3.6$.

We find that the inferred initial shell width, $\Delta_0$, is nearly identical to that obtained in our primary analysis, in which the X-ray data were treated with reduced statistical weight. Therefore, the main conclusion of this paper regarding the finite shell thickness 
remains unchanged and does not depend on the adopted treatment of the X-ray uncertainties.
On the other hand, the fit obtained with the original observational uncertainties requires a significantly larger isotropic-equivalent kinetic energy. The resulting collimation-corrected energy becomes extremely large, implying an implausible jet energetics. 
Although the fit reproduces the X-ray flux more closely, the inferred physical parameters are less realistic. This supports our interpretation that the X-ray flux should be regarded as a weaker constraint on the forward-shock component, possibly owing to additional X-ray emission components and/or systematic uncertainties associated with the cooling-break treatment.

\begin{table}[t]
  \centering
  \renewcommand{\arraystretch}{1.22}
  \caption{The same as Table~1, but for XRF 080330 obtained using the original observational uncertainties.}
  \begin{tabular}{lccc}
    \hline
    Parameters & Prior Ranges & 1D dist & MAP\\
    \hline
    $\log(E_0~[\mathrm{erg}])$      & $[52,56]$      & $55.42^{+0.50}_{-0.36}$ & $55.57$ \\
    $\log\Gamma_0$                 & $[1,4]$        & $3.75^{+0.24}_{-0.43}$  & $3.56$ \\
    $\log(\Delta_0/c~[\mathrm{s}])$& $[1,3]$        & $2.46^{+0.01}_{-0.01}$  & $2.45$ \\
    $\log(n_0~[\mathrm{cm^{k-3}}])$& $[-2,2]$       & $-0.11^{+0.28}_{-0.49}$ & $0.17$ \\
    $k$                            & $[0,2]$        & $0.74^{+0.02}_{-0.02}$  & $0.73$ \\
    $p$                            & $[2.01,2.99]$  & $2.33^{+0.01}_{-0.02}$  & $2.32$ \\
    $\log\epsilon_e$               & $[-3,-0.1]$    & $-2.11^{+0.40}_{-0.56}$ & $-2.25$ \\
    $\log\epsilon_B$               & $[-6,-1]$      & $-5.74^{+0.63}_{-0.24}$ & $-5.97$ \\
    $\log f_\mathrm{e}$            & $[-1,0]$       & $-0.22^{+0.20}_{-0.49}$ & $-0.39$ \\
    $\theta_\mathrm{j}~[\mathrm{rad}]$ & $[0.001,1.5]$ & $0.54^{+0.90}_{-0.44}$ & $0.10$ \\
    $\beta=\theta_{\mathrm{obs}}/\theta_\mathrm{j}$ & $[0,2]$ & $0.64^{+0.31}_{-0.60}$ & $0.68$ \\
    \hline
  \end{tabular}
  \label{table:result_XRF080330_original}
\end{table}

\bibliographystyle{elsarticle-harv} 
\bibliography{example}






\end{document}